\documentclass[lettersize,journal]{IEEEtran} %,draftclsnofoot
\usepackage{amsmath,amsfonts}
\usepackage{algorithmic}
\usepackage{algorithm}
\usepackage{array}
\usepackage[caption=false,font=normalsize,labelfont=sf,textfont=sf]{subfig}
\usepackage{textcomp}
\usepackage{url}
\usepackage{verbatim}
\usepackage{graphicx}
\usepackage{cite}
\usepackage{amssymb}
\usepackage{stfloats}
\usepackage{cuted}
\usepackage{fancyhdr}
\begin{document}

\title{Robust Transmission Design for RIS-Assisted Integrated Sensing and Communication Systems}

\author{Yongqing Xu, \emph{Graduate Student Member, IEEE}, Yong Li, \emph{Member, IEEE}, and Tony Q.S. Quek, \emph{Fellow, IEEE}\\
        
        % <-this % stops a space
        % \thanks{\emph{(Corresponding author: Yong Li)}}% <-this % stops a space
        \thanks{The work of Y. Xu and Y. Li was supported by the National Research and Development Program ``Distributed Large Dimensional Wireless Cooperative Transmission Technology Research and System Verification'' under Grant 2022YFB2902400. The work of Tony Q. S. Quek was supported by the National Research Foundation, Singapore and Infocomm Media Development Authority under its Future Communications Research \& Development Programme. (Corresponding Author: Yong Li)
        	
        	Y. Xu and Y. Li are with the Key Laboratory of Universal Wireless Communications, Beijing University of Posts and Telecommunications, Beijing 100876, China. Email: \{xuyongqing; liyong\}@bupt.edu.cn. 
        	
        	T. Q. S. Quek is with the Singapore University of Technology and Design, Singapore 487372 (e-mail: tonyquek@sutd.edu.sg).}}%

\maketitle
\thispagestyle{fancy}
\cfoot{\small{Copyright \copyright 20xx IEEE. Personal use of this material is permitted. However, permission to use this material for any other purposes must be obtained from the IEEE by sending an email to pubs-permissions@ieee.org.}}
\renewcommand{\headrulewidth}{0mm}

\begin{abstract}
As a critical technology for next-generation communication networks, integrated sensing and communication (ISAC) aims to achieve the harmonious coexistence of communication and sensing. The degrees-of-freedom (DoF) of ISAC is limited due to multiple performance metrics used for communication and sensing. Reconfigurable Intelligent Surfaces (RIS) composed of metamaterials can enhance the DoF in the spatial domain of ISAC systems. However, the availability of perfect Channel State Information (CSI) is a prerequisite for the gain brought by RIS, which is not realistic in practical environments. Therefore, under the imperfect CSI condition, we propose a decomposition-based large deviation inequality approach to eliminate the impact of CSI error on communication rate and sensing Cram\'er-Rao bound (CRB). Then, an alternating optimization (AO) algorithm based on semi-definite relaxation (SDR) and gradient extrapolated majorization-maximization (GEMM) is proposed to solve the transmit beamforming and discrete RIS beamforming problems. We also analyze the complexity and convergence of the proposed algorithm. Simulation results show that the proposed algorithms can effectively eliminate the influence of CSI error and have good convergence performance. Notably, when CSI error exists, the gain brought by RIS will decrease with the increase of the number of RIS elements.
\end{abstract}
%233words


\section{Introduction}
\IEEEPARstart{N}{umerous} emerging applications for future networks, including vehicle-to-everything (V2X), smart homes, and industrial Internet of thing (IoT), require high-quality communication performance and demand accurate sensing capabilities \cite{ref1, ref2}. The communication and sensing functionality can be simultaneously realized in a unified platform due to their similar hardware architecture and signal processing algorithms. Motivated by these considerations, research on integrated sensing and communication (ISAC) is currently underway to empower next-generation communication systems with sensing capabilities, such as mobile edge computing (MEC) \cite{ref3}, federated learning \cite{ref4}, and beamforming \cite{ref5}. ISAC can realize dynamic spectrum sharing between sensing and communication, achieving harmonious co-existence between radar and communication systems \cite{ref6}. 

Reconfigurable intelligent surface (RIS) is a planar surface consisting of a large number of low-cost and nearly passive reconfigurable elements that can be dynamically controlled to induce appropriate amplitude and phase shifts to incident signals \cite{ref7, ref8, ref9, ref10}. RIS has been identified as a promising technology to enhance the energy and spectral efficiency of communication systems and radar systems. Research on deploying RISs into communication systems mainly focuses on joint beamforming design \cite{ref11,ref12,ref13}, robust transmission design \cite{ref14, ref15}, and energy-efficient design \cite{refadd1}. On the other hand, research on deploying RISs into radar systems focuses on enhancing sensing coverage in environments characterized by rich scattering, such as joint beamforming works \cite{ref16, ref17, ref18}, sensing architecture \cite{ref19}, and cellular sensing scheme via multiple RISs \cite{refadd2}. Additionally, RISs can be deployed in scenarios where infrastructures, such as the Global Positioning System (GPS), are insufficient to estimate the locations of UEs and scatterers. To be specific, theoretical bounds for localization have been investigated in \cite{ref20}, signal strength-based localization schemes have been proposed in \cite{ref21}, and time-of-arrival (TOA) and angle of departure (AOD) based methods have been investigated in \cite{ref22,ref23}. 

More important, research on deploying RISs into the ISAC systems mainly focuses on improving communication and sensing performance while eliminating the interference between communication and radar functions. For example, the authors of \cite{ref22} demonstrated that RISs can either increase or maintain the data rate and accurately localize and track user equipment (UE). The authors of \cite{ref24} proposed using RISs to enhance the detection performance of ISAC systems in crowded areas while satisfying the communication signal-noise ratio (SNR) constraints. The authors of \cite{ref25} pointed out that the degrees-of-freedom (DoF) of transmit beamforming is limited when synthesizing desired transmit beam pattern and propose using RISs to minimize multi-user interference. The authors of \cite{ref26} formulated a waveform design problem to maximize the achievable sum rate under beampattern similarity constraints. Moreover, the authors of \cite{ref27} proposed using RISs to minimize interference between communication and radar functions, while in \cite{ref28}, the authors proposed deploying two RISs near the transmitter and receiver to enhance communication signals and suppress mutual interference. 

In general, obtaining accurate channel state information (CSI) of the base station (BS)-RISs, RISs-UEs links, and BSs-UEs links, as well as reflection coefficients of radar targets, is crucial for optimizing amplitude and phase shifts of RISs \cite{ref29}. Various methods have been proposed to obtain CSI, such as discrete Fourier transform (DFT)-based methods \cite{ref30}, compressed sensing (CS)-based channel estimation schemes \cite{ref31,ref32,ref33}, subspace-based estimation methods \cite{ref34,ref35,ref36}, and channel estimation algorithms based on machine learning \cite{ref37}. However, the performance of RIS-assisted systems largely depends on the accuracy of the CSI and radar coefficients. Most existing works on joint BS and RIS beamforming design assume that the CSI and radar coefficients have been perfectly estimated. Unfortunately, the channels are difficult to estimate due to the passive feature of RIS and the dynamic environments. To address this, the authors of \cite{ref14} designed two robust transmission schemes for worst-case CSI error case and statistic CSI error case, using the Bernstein-type inequality \cite{ref39} to approximate the outage probability constraint; the authors of \cite{ref40} solved the robust beamforming problem under worst-case CSI error constraints and discrete phase shifts constraints on RIS, using a simple mapping method to map continuous RIS phase shifts to discrete ones; the authors of \cite{ref41, ref42, ref43} aimed to minimize BS transmit power under the outage probability constraint. Additionally, the authors of \cite{ref44} used the brand-and-bound (BAB) method to handle the non-convex beamforming problem induced by RIS phase shifts. However, the existing approximation method \cite{ref14} for dealing with the outage probability has a high complexity, the mapping operation for handling the discrete RIS phase shifts \cite{ref40} introduces additional errors, and the BAB method \cite{ref44} incurs a high computational burden when the dimensions of variables are large.

\subsection{The Contribution of This Work}
Our study focuses on robust beamforming for RIS-assisted ISAC systems. We propose a novel approximation method to approximate the probability constraints for the communication rate and the lower bound of the estimation accuracy, i.e., the Cram\'er-Rao bound (CRB). We then propose an alternating optimization (AO) algorithm to solve the robust beamforming problem, with two schemes for handling the non-convex discrete RIS phase shifts constraints. Our proposed robust beamforming method is demonstrated to approach the probability constraints tightly and handle the discrete RIS phase shifts effectively.

The main contributions of our work are:
\begin{itemize}
	\item We establish a robust transmission model for RIS-assisted ISAC systems, incorporating two signal models for communication and sensing functions, respectively.
	\item We propose a novel approximation method based on decomposition-based large deviation inequality that only involves simple second-order constraints (SOC) to accurately approximate the probability constraints for the communication rate and CRB.
	\item We propose an AO algorithm that transforms the transmit beamforming into a semi-definite programming (SDP) problem using the semi-definite relaxation (SDR) method. We solve the RIS beamforming constrained by the discrete RIS phase shifts using two schemes. The first scheme relaxes the discrete RIS phase shifts as continuous ones. The second scheme transforms the discrete constraint into a square penalty (SP) term. 
	\item We propose an accelerated projected gradient (APG) based gradient extrapolated majorization-maximization (GEMM) method for efficiently solving the SP problem.
	\item Simulation results are manifested to verify the effectiveness of the proposed AO algorithm and the approximation method. The computational complexity of the proposed algorithms is also analyzed.
\end{itemize}
\subsection{Paper Organization}
The remainder of this paper is organized as follows. Section \uppercase\expandafter{\romannumeral2} introduces the system model and the signal models. Section \uppercase\expandafter{\romannumeral3} formulates the optimization model for minimizing the transmit power under the probability constraints and introduces the approximation methods. Section \uppercase\expandafter{\romannumeral3} also introduces the AO algorithm for solving the robust beamforming problem. Section \uppercase\expandafter{\romannumeral4} studies the complexity and convergence of the proposed algorithms. Section \uppercase\expandafter{\romannumeral5} presents the simulation results. Section \uppercase\expandafter{\romannumeral6} concludes the paper.

\subsection{Notations}
Throughout this paper, $\boldsymbol{S}$ denotes a matrix, $\boldsymbol{s}$ denotes a vector, $s$ denotes a scalar. $[\cdot]^T$ and $[\cdot]^H$ stand for the transpose and complex conjugate operations. $\mathcal{CN}$ denotes the circularly-symmetric Gaussian random distribution. $\mathbb{C}^{\left(\cdot\right)}$ denotes the complex space. $\mathbb{E}\left[\cdot\right]$ denotes the statistical expectation. $\text{vec}(\boldsymbol{s})$ and $\text{vec}(\boldsymbol{S})$ represents two vectors which are composed of $\boldsymbol{s}$ and $\boldsymbol{S}$. $\text{diag}(\boldsymbol{s})$ represents a matrix whose diagonal entries are composed of $\boldsymbol{s}$. $\text{diag}(\boldsymbol{S})$ represents a vector whose entries are composed of the diagonal entries of $\boldsymbol{S}$. $\text{Tr}(\cdot)$ represents the trace of a matrix. $\text{Pr}(\cdot)$ represents the cumulative distribution function. $\Vert\cdot\Vert_F$ stands for the Frobenius norm. $\Vert\cdot\Vert$ stands for the $l_2$ norm. $\textbf{I}$stands for the identify matrix. $\text{Re}(\cdot)$ and $\text{Im}(\cdot)$ stand for the real component and the imaginary component of the argument. $\otimes$ stands for the Kronecker product. $\text{rank}(\cdot)$ stands for the rank of a matrix. $\left(\boldsymbol{S}\right)_{k,k}$ stands for the $k$-th diagonal element. $\nabla_{\boldsymbol{s}}f(\boldsymbol{s})$ stands for the derivation of $f$. $\left\langle\cdot,\cdot\right\rangle$ stands for the Euclidean inner product. $\left[x\right]_a^b$ denotes $\text{min}\{b,\text{max}\{x,a\}\}$. $\lfloor x \rfloor$. $\lfloor s \rfloor$ and $\angle s$ denote rounding towards minus infinity and the argument of $s$, respectively.

\section{System Model}
\begin{figure}[!t]
        \centering
        \includegraphics[width=0.45\textwidth]{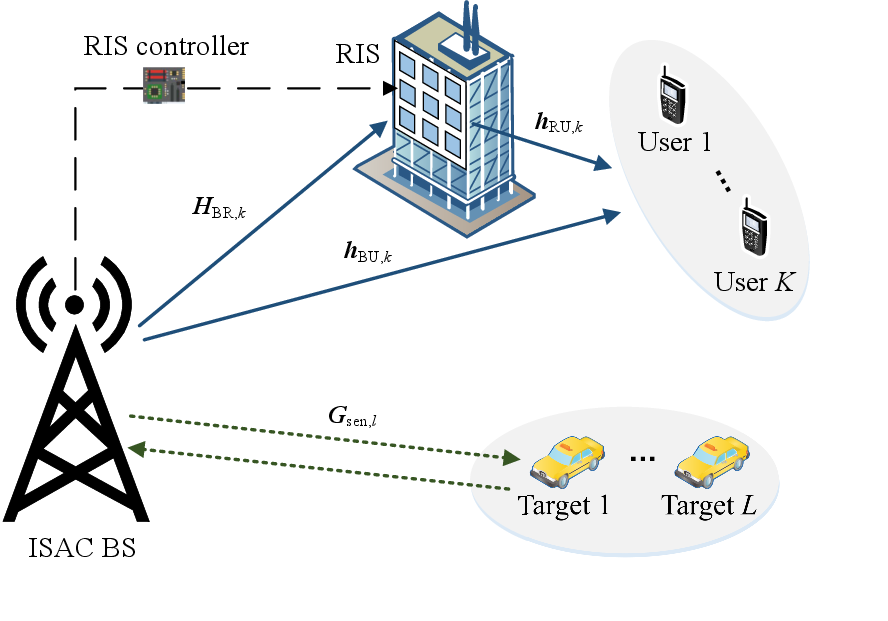}
        \caption{A RIS-assisted ISAC system.}
        \label{fig_systemmodel}
\end{figure}

The considered system is illustrated in Fig. 1, where an integrated sensing and communication (ISAC) BS with $N$ transmit antennas and $N_{\text{R}}$ receive antennas communicates with $K$ single-antenna user while sensing $L$ targets\footnote{As illustrated by \cite{ref19}, the performance gain of the CRB of angle estimation from the reflected signal path of the passive RIS is minimal when the line-of-sight (LoS) signal paths between the BS and the targets exist. Therefore, we neglect the role of RIS in sensing. The robust transmission in non-line-of-sight (NLoS) scenarios will be discussed in our future work.}, with the aid of an $M$-element RIS. The transmit antennas and receive antennas are both uniform linear arrays (ULAs). The transmit and receive antennas at the BS are widely separated to suppress the leakage signals between them. We assume that the inter-distance of the transmit antennas, the receive antennas, and RIS elements are half-wavelength. The set of RIS elements, communication users, and targets are denoted by $ \mathcal{M} $, $ \mathcal{K} $, and $ \mathcal{L} $. The phase shifts of the RIS elements are controlled by a RIS controller, which communicates wirelessly with the ISAC BS for cooperative transmission. The phase matrix of the RIS is represented by $\boldsymbol{\Theta}\triangleq\text{diag}\left(\theta_1,\cdots,\theta_M\right)$. We assume an ideal RIS reflection magnitude model where each element is fixed at $\vert\theta_m\vert=1,\forall m\in\mathcal{M}$, as well as a practical RIS reflection phase shifts model where each $\theta_m$ can only take $d$ finite values, equally spaced in $[0, 2\pi)$. The discrete set of phase shifts for each RIS element is given by 
\begin{equation} 
        \label{eq1}
        \mathcal{F} \triangleq \{\theta | \theta = e^{(j\frac{2\pi}{d}i + \frac{\pi}{d})}, 
        i = 0, ..., d-1\}. 
\end{equation}

\subsection{Communication Signal Model}
The received signal at the $k$-th communication user can be expressed as
\begin{subequations}    
\begin{align}
        \boldsymbol{y}_{\text{com},k} &= \left(\boldsymbol{h}_{\text{BU},k}^H+\boldsymbol{h}_{\text{RU},k}^H\boldsymbol{\Theta}\boldsymbol{H}_{\text{BR}}^H\right)\boldsymbol{s}_k\boldsymbol{x}_k \nonumber \\
        &+\sum_{i=1,i\neq k}^{K}\left(\boldsymbol{h}_{\text{BU},k}^H+\boldsymbol{h}_{\text{RU},k}^H\boldsymbol{\Theta}\boldsymbol{H}_{\text{BR}}^H\right)\boldsymbol{s}_i\boldsymbol{x}_i +\boldsymbol{z}_k, \label{eq2} \tag{2}
\end{align}
\end{subequations}
where $\boldsymbol{h}_{\text{BU},k}\in \mathbb{C}^{N \times 1}$ is the channel vector between the BS and the $k$-th user, $\boldsymbol{h}_{\text{RU},k}\in \mathbb{C}^{M \times 1}$ is the channel vector between the RIS and the $k$-th user, and $\boldsymbol{H}_{\text{BR}}\in \mathbb{C}^{N \times M}$ is the channel matrix between the BS and the RIS. The transmit beamforming matrix of the ISAC BS is $\boldsymbol{S}=\left[\boldsymbol{s}_1, \boldsymbol{s}_2, \cdots, \boldsymbol{s}_K\right]\in\mathbb{C}^{N\times K}$. The transmit signal matrix is $ \boldsymbol{X}=\left[  \boldsymbol{x}_1, \boldsymbol{x}_2, \cdots, \boldsymbol{x}_K\right]^T \in \mathbb{C}^{K\times T}$, where $T$ is the number of signal samples. The signal vectors in $\boldsymbol{X}$ are assumed to be statistically independent, i.e., $\mathbb{E}\left[\boldsymbol{X}\boldsymbol{X}^H\right]=\textbf{I}_K$. Moreover, $\boldsymbol{z}_k$ is the additive white Gaussian noise (AWGN) at the $k$-th user, which is assumed to follow the circularly symmetric complex Gaussian (CSCG) distribution, i.e., $\boldsymbol{z}_k\sim\mathcal{CN}\left(0,\sigma_{\text{com},k}^2\textbf{I}\right)$ with $\sigma_{\text{com},k}^2$ denoting the variance of the AWGN.

Due to the passive characteristic of the RIS, the cascaded BS-RIS-user channel at the transmitter (CBRUT) is difficult to obtain. The direct channel state information at the transmitter (DCSIT) is also difficult to obtain in the complex wireless propagation environments. Hence, we assume the imperfect DCSIT and the imperfect CBRUT\footnote{The channel estimation method in \cite{ref34} can be used to estimated the imperfect DCSIT and the imperfect CBRUT.}. The imperfect DBIUT can be expressed as
\begin{equation} 
        \label{eq3}
        \boldsymbol{h}_{\text{BU},k} =  \hat{\boldsymbol{h}}_{\text{BU},k} + \Delta\boldsymbol{h}_{\text{BU},k}, \forall k \in \mathcal{K}, 
\end{equation}
where $\hat{\boldsymbol{h}}_{\text{BU},k}$ is the estimated channel state information (CSI) known at the BS, and $\Delta\boldsymbol{h}_{\text{BU},k}$ is the unknown CSI error. Similarly, the imperfect CBRUT is expressed as
\begin{equation} 
        \label{eq4}
        \boldsymbol{H}_{\text{BRU},k} =  \hat{\boldsymbol{H}}_{\text{BRU},k} + \Delta\boldsymbol{H}_{\text{BRU},k}, \forall k \in \mathcal{K}, 
\end{equation}
where $\boldsymbol{H}_{\text{BRU},k}\triangleq\text{diag}\left(\boldsymbol{h}_{\text{RU},k}\right)^H\boldsymbol{H}_{\text{BR}}^H\in\mathbb{C}^{M\times N}$ is the cascaded BS-RIS-user channel. The CSI error, i.e., $\Delta\boldsymbol{h}_{\text{BU},k}$ and $\Delta\boldsymbol{H}_{\text{BRU},k}$ are all assumed to follow the CSCG distribution\footnote{According to \cite{ref34}, the variances of the CSI error are only related to the transmit power of the signal during channel estimation, the received combining vectors, the covariance matrices of the received signals, and the noise variance. All of these parameters are known, so we assume that the variances of the CSI error are also known.}. The CSI error can be expressed as
\begin{equation} 
        \label{eq5}
        \text{vec}\left(\Delta\boldsymbol{h}_{\text{BU},k}\right) \sim \mathcal{CN}\left(0, \boldsymbol{\Xi}_{\text{BU},k}\right), \forall k \in \mathcal{K}, 
\end{equation}
\begin{equation} 
        \label{eq6}
        \text{vec}\left(\Delta\boldsymbol{H}_{\text{BRU},k}\right) \sim \mathcal{CN}\left(0, \boldsymbol{\Xi}_{\text{BRU},k}\right), \forall k \in \mathcal{K}, 
\end{equation}
where $\boldsymbol{\Xi}_{\text{BU},k}\in \mathbb{C}^{N\times N}$ and $\boldsymbol{\Xi}_{\text{BRU},k}\in \mathbb{C}^{MN\times MN}$ are semi-definite error covariance matrices. 

Further, the average interference plus noise for the $k$-th user can be written from (\ref{eq2}), (\ref{eq3}), and (\ref{eq4}) as       
\begin{equation}
                \label{eq7}
                \zeta_k = \sum_{i=1,i\neq k}^{K}\left\vert\left(\hat{\boldsymbol{h}}_{\text{BU},k}^H+\boldsymbol{\theta}^H\hat{\boldsymbol{H}}_{\text{BRU},k}\right)\boldsymbol{s}_i\right\vert^2+\sigma_{\text{com},k}^2,
\end{equation}
where $\boldsymbol{\theta}= \text{diag}\left(\boldsymbol{\Theta}\right)$. Then, the communication rate for the $k$-th user can be expressed as
\begin{equation}
        \label{eq8}
        R_{k}=\log_2 \left(1+\frac{1}{\zeta_{k}}\left\vert\left(\hat{\boldsymbol{h}}_{\text{BU},k}^H+\boldsymbol{\theta}^H\hat{\boldsymbol{H}}_{\text{BRU},k}\right)\boldsymbol{s}_k\right\vert^{2}\right).
\end{equation}

\subsection{Sensing Signal Model}
The received sensing signal at the ISAC BS is $\boldsymbol{Y}_{\text{sen}}$, which can be expressed as
\begin{subequations}
        \label{eq9}
\begin{align}
        \boldsymbol{Y}_{\text{sen}} &= \sum_{l=1}^{L}\left(\boldsymbol{G}_{\text{sen},l}\boldsymbol{SX}\right)+\boldsymbol{W} \nonumber\\ 
        &= \boldsymbol{A\Sigma A}^T\boldsymbol{SX} + \boldsymbol{W},\tag{9}
\end{align}
\end{subequations} 
where $\boldsymbol{G}_{\text{sen},l} \triangleq \alpha_l\boldsymbol{\mathrm{b}}\left(\varphi_l\right)\boldsymbol{\mathrm{a}}\left(\phi_l\right)^T\in \mathbb{C}^{N \times N}$ denotes the sensing response matrix, $\alpha_l$ is the reflection coefficient of the $l$-th target, including the propagation loss and the radar cross section (RCS), $\phi_l$ and $\varphi_l$ are respectively the angle of departure (AoD) and the angle of arrival (AoA) of the $l$-th target, $\boldsymbol{\mathrm{a}}\left(\phi\right) \triangleq \left[1,e^{j\pi \sin(\phi)},\dots,e^{j\pi(N-1)\sin(\phi)}\right]^T \in \mathbb{C}^{N \times 1}$ is the steering vector of the ISAC transmit antennas, $\boldsymbol{\mathrm{b}}\left(\varphi\right) \triangleq \left[1,e^{j\pi \sin(\varphi)},\dots,e^{j\pi(N_{\text{R}}-1)\sin(\varphi)}\right]^T \in \mathbb{C}^{N_{\text{R}} \times 1}$ is the steering vector of the ISAC receive antennas, and $\boldsymbol{W}$ is the AWGN, which is assumed to follow the CSCG distribution, i.e., $\boldsymbol{W}\sim \mathcal{CN}\left(0,\sigma_{\text{sen}}^2\textbf{I}\right)$. Moreover, we define $\boldsymbol{A}\triangleq\left[\boldsymbol{\mathrm{a}}\left(\phi_1\right),\cdots,\boldsymbol{\mathrm{a}}\left(\phi_L\right)\right]$, $\boldsymbol{B}\triangleq\left[\boldsymbol{\mathrm{b}}\left(\varphi_1\right),\cdots,\boldsymbol{\mathrm{b}}\left(\varphi_L\right)\right]$, and $\boldsymbol{\Sigma}\triangleq\text{diag}\left(\alpha_1,\cdots,\alpha_L\right)$. 

It is noted that in Eq. (\ref{eq9}), the echoes from $L$ targets are mixed together. We can use the MUSIC algorithm proposed in \cite{refadd3} to estimate the angles of multiple targets. This enables the feasibility of separately optimizing the CRB of angle estimation for each target. Additionally, we focus on the optimal achievable performance of angle estimation without concerning specific angle estimation algorithms.

The CRB is a lower bound on the variance of any unbiased parameter estimator. The CRB for the direction of arrival (DoA) of the $l$-th target is defined as
\begin{equation}
        \label{eq10}
        \text{CRB}_l = \frac{\sigma_{\text{sen}}^2}{2\vert\alpha_l\vert^2}\left[\text{Tr}\left(\boldsymbol{S}^H\dot{\boldsymbol{A}}_l^H\dot{\boldsymbol{A}}_l\boldsymbol{S}\right)\right]^{-1},
\end{equation}
where $\dot{\boldsymbol{A}}_l\triangleq\dot{\boldsymbol{\mathrm{b}}}\left(\varphi_l\right)\boldsymbol{\mathrm{a}}\left(\phi_l\right)^T+\boldsymbol{\mathrm{b}}\left(\varphi_l\right)\dot{\boldsymbol{\mathrm{a}}}\left(\phi_l\right)^T$ with $\dot{\boldsymbol{\mathrm{a}}}\left(\phi_l\right)$ and $\dot{\boldsymbol{\mathrm{b}}}$ denoting the derivatives of the transmit steering vector and the receive steering vector, and they can be expressed as
\begin{equation}
	\label{eq11}
	\begin{aligned}	&\dot{\boldsymbol{\mathrm{a}}}\left(\phi\right)=\left[0,j\pi\text{cos}(\phi)\mathrm{a}(2),\cdots,j\pi(N-1)\text{cos}(\phi)\mathrm{a}(N)\right]^T,\\
		&\dot{\boldsymbol{\mathrm{b}}}\left(\varphi\right)=\left[0,j\pi\text{cos}(\varphi)\mathrm{b}(2),\cdots,j\pi(N_{\text{R}}-1)\text{cos}(\varphi)\mathrm{b}(N_{\text{R}})\right]^T,
	\end{aligned}
\end{equation}
where $\mathrm{a}(n)$ and $\mathrm{b}(n)$ denote the $n$-th entry of $\boldsymbol{\mathrm{a}}\left(\phi\right)$ and $\boldsymbol{\mathrm{b}}\left(\phi\right)$, respectively. The CRB are derived in Appendix A.

The accurate RCS and reflection coefficient are difficult to obtain due to the clutter and the mobility of the targets in environments characterized by rich scattering. We assume the imperfect reflection coefficients at the transmitter (RCT), which can be expressed as
\begin{equation}
        \label{eq12}
        \alpha_l=\hat{\alpha}_l+\Delta\alpha_l,\forall l\in \mathcal{L},
\end{equation}
where $\hat{\alpha}_l$ is the estimated reflection coefficient, and $\Delta\alpha_1$ is the unknown RCT error, which is assume to follow the CSCG distribution. The RCT error can be written as
\begin{equation}
        \label{eq13}
        \Delta\alpha_l\sim\mathcal{CN}(0,\varepsilon_l^2),\forall l\in \mathcal{L},
\end{equation}
where $\varepsilon_l^2$ is the variance of the RCT error for the $l$-th target.

\noindent \textbf{Remark 1.} \emph{When the CSI and RCT errors are minor, it is possible to design effective and robust transmission schemes that meet the requirements for communication rate and sensing CRB. However, when the CSI error and RCT error are relatively large, effective robust transmission methods do not exist due to the randomness of the CSI and reflection coefficients. In addition, we can think the DoA estimation is invalid when the CRB is large.}

\noindent \textbf{Remark 2.} \emph{The value of the CRB in (\ref{eq10}) depends on the angle values. On one hand, we can minimize the CRB in the direction of interest, where the potential targets may exist. On the other hand, when the target is static or moves slowly, it is not necessary to update the beamforming matrices frequently, and the estimated angles previously are sufficient to characterize the CRB of angle estimation. Therefore, we assume that the angles of the targets are known.}

\section{Outage Probability Constrained Robust Beamforming}
In this section, we first formulate a robust beamforming optimization problem with two probability constraints and one discrete RIS phase shifts constraint. Subsequently, we propose an approximation method to approximate the two probability constraints safely. Finally, we present an AO algorithm that solves the coupled robust beamforming problem.

\subsection{Problem Formulation}
The robust beamforming optimization problem for minimizing the transmit power is formulated as
\begin{subequations}
        \label{eq14}
        \begin{align}
                \min_{\boldsymbol{S}, \boldsymbol{\theta}} \quad & \Vert \boldsymbol{S} \Vert_F^2 \label{eq14a} \tag{14a}\\
                \text{s.t.} \quad & \text{Pr}\{R_k \ge r_k\} \ge 1-\rho_k, \forall k \in \mathcal{K},\label{eq14b} \tag{14b} \\
                &\text{Pr}\{\text{CRB}_l\le c_l\} \ge 1-p_l,\forall k \in \mathcal{K},\label{eq14c} \tag{14c} \\
                & \boldsymbol{\theta}_m \in \mathcal{F}, \forall m \in \mathcal{M}, \label{eq14d} \tag{14d}
        \end{align}
\end{subequations}
where (\ref{eq14b}) is the outage probability constraint for the communication rate, $r_k$ is the rate threshold, and $\rho_k$ is the outage probability; (\ref{eq14c}) is the failure probability constraint for the DoA estimation, $c_l$ is the CRB threshold, and $p_k$ is the failure probability; (\ref{eq14d}) is the unit modulus and the discrete phase shifts constraint for the reflection coefficients at the RIS. Note that the probability constraints (\ref{eq14b}) and (\ref{eq14c}) and the discrete constraint (\ref{eq14d}) are non-convex. We first propose two methods to approximate the two probability constraints. Then, we propose an AO algorithm to solve the problem in (\ref{eq14a}), which solves the transmit beamforming subproblem and the RIS beamforming subproblem iteratively.

\subsection{Probability Constraint Approximation}
\subsubsection{Approximation for Communication Rate}
The tail probability in (\ref{eq14b}) can be rewritten as
\begin{subequations}
        \label{eq15}
        \begin{align}
                \text{Pr}&\left\{\log_2\left(1 + \frac{\left|\boldsymbol{c}_k\left(\boldsymbol{\theta}\right)
                        \boldsymbol{s}_k\right|^2}{\left\Vert \boldsymbol{c}_k\left(\boldsymbol{\theta}\right)
                        \boldsymbol{S}_{-k} \right\Vert_2^2 + \sigma_{\text{com},k}^2} \right) \ge r_k  \right\} \nonumber\\
                    &= \text{Pr}\left\{\boldsymbol{c}_k\left(\boldsymbol{\theta}\right)\left[\boldsymbol{s}_k\boldsymbol{s}_k^H-\left(2^{r_k}-1\right)\boldsymbol{S}_{-k}\boldsymbol{S}_{-k}^H\right]\boldsymbol{c}_k^H\left(\boldsymbol{\theta}\right)\ge 0\right\}\nonumber\\
                &=\text{Pr}\left\{ \boldsymbol{c}_k\left(\boldsymbol{\theta}\right) \boldsymbol{\Psi}_k \boldsymbol{c}_k^H\left(\boldsymbol{\theta}\right) - \sigma_{\text{com},k}^2 \ge 0\right\},\tag{15}
        \end{align}
\end{subequations}
where $\boldsymbol{c}_k\left(\boldsymbol{\theta}\right)=\boldsymbol{h}_{\text{BR},k}^H + \boldsymbol{\theta}^H\boldsymbol{H}_{\text{BRU},k}$, $\boldsymbol{S}_{-k}\triangleq\left(\boldsymbol{s}_1,\cdots,\boldsymbol{s}_{k-1},\boldsymbol{s}_{k+1},\cdots,\boldsymbol{s}_K\right)$, and $ \boldsymbol{\Psi}_k =\left(2^{r_k}-1\right)^{-1}\boldsymbol{s}_k\boldsymbol{s}_k^H-\boldsymbol{S}_{-k}\boldsymbol{S}_{-k}^H$.

In order to approximate the tail probability (\ref{eq15}), we propose the following lemma (Lemma 1).

\noindent \textbf{Lemma 1.} \emph{(Decomposition-Based Large Deviation Inequality \cite{ref39})} \emph{Let} $\boldsymbol{x} \sim \mathcal{CN} \left( 0,\textbf{I}_n \right), $ \emph{and} $ \boldsymbol{Q} \in \mathbb{H}^{n \times n}, $ \emph{and} $ \boldsymbol{r} \in \mathbb{C}^{n \times 1} $ \emph{be given. Then, for any} $ v > 1/\sqrt{2} $ \emph{and} $ \eta > 0, $ \emph{we have}
\begin{subequations}
        \label{eq16}
        \begin{align}
                &\text{Pr}\{\boldsymbol{x}^H\boldsymbol{Q}\boldsymbol{x}+2\text{Re}\{\boldsymbol{r}^H\boldsymbol{x}\} \le \text{Tr}\{\boldsymbol{Q}\} - \eta\} \nonumber \\
                \le &\left\{
                \begin{array}{lr}
                        \text{exp} \left( -\frac{\eta^2}{4T^2} \right) &\text{for } 0 < \eta \le 2 \bar{\theta}vT,\\
                        \text{exp} \left( -\frac{\bar{\theta}v\eta}{T} + \left( \bar{\theta}v \right)^2 \right) &\text{for } \eta > 2 \bar{\theta}vT,
                \end{array}
                \right.\tag{16}
        \end{align}
\end{subequations}
\emph{where} $ \bar{\theta} = 1-1/2v^2$, $ T = v \Vert \boldsymbol{Q} \Vert_F + \frac{1}{\sqrt{2}} \Vert \textbf{r} \Vert$. 

\emph{Proof: }Please refer to Lemma 2 in \cite{ref39}.$\hfill\blacksquare$

Next, we use Lemma 1 to approximate (\ref{eq15}). First, we define $\boldsymbol{\Xi}_{\text{BU},k}\triangleq\gamma_{\text{BU},k}^2\textbf{I}$ and $\boldsymbol{\Xi}_{\text{BRU},k}\triangleq\gamma_{\text{BRU},k}^2\textbf{I}$. We have $\Delta\boldsymbol{h}_{\text{BU},k}=\gamma_{\text{BU},k}\boldsymbol{e}_{\text{BU},k}$ and $\Delta\boldsymbol{H}_{\text{BRU},k}=\gamma_{\text{BRU},k}\boldsymbol{E}_{\text{BRU},k}$, where $\boldsymbol{e}_{\text{BU},k}\sim\mathcal{CN}(0,\textbf{I})$ and $\boldsymbol{E}_{\text{BRU},k}\sim\mathcal{CN}(0,\textbf{I})$. Then, (\ref{eq15}) is reformulated as the following equation by substituting (\ref{eq5}) and (\ref{eq6}) into (\ref{eq15}), which is written as
\begin{subequations}
\label{eq17}
\begin{align}
        &\text{Pr}\left\{ \left[ \left(\hat{\boldsymbol{h}}_{\text{BU},k}^H+\Delta\boldsymbol{h}_{\text{BU},k}^H\right)+\boldsymbol{\theta}^H\left(\hat{\boldsymbol{H}}_{\text{BRU},k}+\Delta\boldsymbol{H}_{\text{BRU},k}\right) \right]\right. \nonumber \\
        &\enspace\enspace\boldsymbol{\Psi}_k\left[ \left(\hat{\boldsymbol{h}}_{\text{BU},k}+\Delta\boldsymbol{h}_{\text{BU},k}\right)+\left(\hat{\boldsymbol{H}}_{\text{BRU},k}^H+\Delta\boldsymbol{H}_{\text{BRU},k}^H\right)\boldsymbol{\theta} \right] \nonumber \\
        &\enspace\enspace-\sigma_{\text{com},k}^2 \ge 0\Big\} \nonumber \\
        &=\text{Pr}\left\{\left(\hat{\boldsymbol{h}}_{\text{BU},k}^H+\boldsymbol{\theta}^H\hat{\boldsymbol{H}}_{\text{BRU},k}\right)\boldsymbol{\Psi}_k\left(\hat{\boldsymbol{h}}_{\text{BU},k}+\hat{\boldsymbol{H}}_{\text{BRU},k}^H\boldsymbol{\theta}\right)+\right.\nonumber \\
        &2\text{Re}\left[\left(\hat{\boldsymbol{h}}_{\text{BU},k}^H+\boldsymbol{\theta}^H\hat{\boldsymbol{H}}_{\text{BRU},k}\right)\boldsymbol{\Psi}_k\left(\Delta\boldsymbol{h}_{\text{BU},k}+\Delta\boldsymbol{H}_{\text{BRU},k}^H\boldsymbol{\theta}\right)\right] \nonumber \\
        &+\left(\Delta\boldsymbol{h}_{\text{BU},k}^H+\boldsymbol{\theta}^H\Delta\boldsymbol{H}_{\text{BRU},k}\right)\boldsymbol{\Psi}_k\left(\Delta\boldsymbol{h}_{\text{BU},k}+\Delta\boldsymbol{H}_{\text{BRU},k}^H\boldsymbol{\theta}\right) \nonumber \\
        &\enspace\enspace-\sigma_{\text{com},k}^2\ge 0 \Big\}. \tag{17}
\end{align}
\end{subequations}
The outage probability constraint (\ref{eq14b}) is formulated and approximated in the following propositions (Proposition 1 and Proposition 2).

\noindent \textbf{Proposition 1.} \emph{The outage probability constraint can be rewritten as}
\begin{subequations}
        \label{eq18}
        \begin{align}
                &\text{Pr}\left\{\boldsymbol{e}_k^H\boldsymbol{Q}_k\boldsymbol{e}_k+2\text{Re}\left(\boldsymbol{r}_k^H\boldsymbol{e}_k\right) + s_k \ge 0 \right\} \nonumber \\
                &\quad \ge 1-\rho_k, 
                \forall k \in \mathcal{K},      \tag{18}
        \end{align}
\end{subequations}
\emph{where}
\begin{equation}
        \label{eq19}
        \boldsymbol{e}_k=\left[\boldsymbol{e}_{\text{BU},k}^H,\text{vec}\left(\boldsymbol{E}_{\text{BRU},k}\right)^T\right]^H,
\end{equation}
\begin{equation}
        \label{eq20}
        \setlength{\arraycolsep}{0pt}
        \boldsymbol{Q}_{k} = \begin{bmatrix}
                \gamma_{\text{BU},k}^2\boldsymbol{\Psi}_k, & \gamma_{\text{BU},k}\gamma_{\text{BRU},k}\left(\boldsymbol{\Psi}_k\otimes\boldsymbol{\theta}^T\right)\\
                \gamma_{\text{BU},k}\gamma_{\text{BRU},k}\left(\boldsymbol{\Psi}_k\otimes\boldsymbol{\theta}^*\right), & \gamma_{\text{BRU},k}^2\left(\boldsymbol{\Psi}_k\otimes(\boldsymbol{\theta}\boldsymbol{\theta}^H)\right)
        \end{bmatrix},
\end{equation}
\begin{subequations}
        \label{eq21}
        \begin{align}
                \boldsymbol{r}_k &= \bigg[\gamma_{\text{BU},k}\boldsymbol{\Psi}_k\left(\hat{\boldsymbol{h}}_{\text{BU},k}+\hat{\boldsymbol{H}}_{\text{BRU},k}^H\boldsymbol{\theta}\right); \nonumber \\
                &\enspace\enspace\enspace\enspace\gamma_{\text{BRU},k}\text{vec}\left[\boldsymbol{\theta}\left(\hat{\boldsymbol{h}}_{\text{BU},k}^H+\boldsymbol{\theta}^H\hat{\boldsymbol{H}}_{\text{BRU},k}\right)\boldsymbol{\Psi}_k\right]^*\bigg],\tag{21}
        \end{align}
\end{subequations}
\emph{and}
\begin{subequations}
        \label{eq22}
        \begin{align}
                &s_k= \left(\hat{\boldsymbol{h}}_{\text{BU},k}^H+\boldsymbol{\theta}^H\hat{\boldsymbol{H}}_{\text{BRU},k}\right)\boldsymbol{\Psi}_k\left(\hat{\boldsymbol{h}}_{\text{BU},k}+\hat{\boldsymbol{H}}_{\text{BRU},k}^H\boldsymbol{\theta}\right)\nonumber\\
                &\quad\quad-\sigma_{\text{com},k}^2.\tag{22}
        \end{align}
\end{subequations}

\emph{Proof: }The proof is given in Appendix B.$\hfill\blacksquare$

\noindent \textbf{Proposition 2.} \emph{The outage probability constraint (\ref{eq18}) can be safely approximated by Lemma 1 into the following SOC constraint}
\begin{equation}
        \label{eq23}
        \mathrm{Tr}\{\boldsymbol{Q}_k\} + s_k \ge 2\sqrt{\text{ln}(1/\rho_k)}T_k, \forall k \in \mathcal{K},
\end{equation}
\emph{where} $T_k=v_k\Vert\boldsymbol{Q}_k\Vert_F+\frac{1}{\sqrt{2}}\Vert\boldsymbol{r}_k\Vert$, \emph{and} $v_k$ \emph{is the solution of the equation} $\left[1-1/\left(2v_k^2\right)\right]v_k=\sqrt{\ln\left(1/\rho_k\right)}$. \emph{The constraint (\ref{eq23}) can be rewritten as}
\begin{equation}
        \label{eq24}
        \mathrm{Tr}\{\boldsymbol{Q}_k\} + s_k \ge 2\sqrt{\text{ln}(1/\rho_k)}(x_k+y_k),\forall k \in \mathcal{K},
\end{equation}
\begin{equation}
        \label{eq25}
        \frac{1}{\sqrt{2}}\Vert\boldsymbol{r}_k\Vert\leq x_k,\forall k \in \mathcal{K},
\end{equation}
\begin{equation}
        \label{eq26}
        v_k\Vert \boldsymbol{Q}_k \Vert_F \le y_k, \forall k \in \mathcal{K},
\end{equation}
\emph{where} $\boldsymbol{x}=\left[x_1,\cdots,x_K\right]^T$ \emph{and} $\boldsymbol{y}=\left[y_1,\cdots,y_K\right]^T$ \emph{are the auxiliary variables}.

\emph{Proof: }The proof is given in Appendix C.$\hfill\blacksquare$

Now, the outage probability constraint (\ref{eq14b}) is transformed into the convex constraints through Proposition 2. The SOC constraints in (\ref{eq24}),(\ref{eq25}), and (\ref{eq26}) are easily dealt with by the convex optimization methods, such as the inner point method \cite{ref45}.

\subsubsection{Approximation for Sensing CRB}
The tail probability in (\ref{eq14c}) can be rewritten as
\begin{equation}
        \label{eq27}
        \text{Pr}\left\{\vert\hat{\alpha}_l+\Delta\alpha_l\vert^2\ge\frac{1}{2c_l\sigma_{\text{sen}}^2\text{Tr}\left(\boldsymbol{S}^H\dot{\boldsymbol{A}}_l^H\dot{\boldsymbol{A}}_l\boldsymbol{S}\right)}\right\}.
\end{equation}
Then, we use Lemma 1 to approximate (\ref{eq27}). We define $\Delta\alpha_l=\varepsilon_le_{\text{sen},l}$, where $e_{\text{sen},l}\sim\mathcal{CN}(0,1)$. The probability constraint (\ref{eq14c}) is reformulated as
\begin{equation}
        \label{eq28}
        \text{Pr}\left\{e_{\text{sen},l}^*\tilde{q}_le_{\text{sen},l}+2\text{Re}\left(\tilde{r}_l^*e_{\text{sen},l}\right)+\tilde{s}_l\ge0\right\}\ge 1-p_l,
\end{equation}
where $\tilde{q}_l=\varepsilon_l^2$, $\tilde{r}_l=\varepsilon_l\hat{\alpha}_l$, and $\tilde{s}_l=\vert\hat{\alpha}_l\vert^2-\left[2c_l\sigma_{\text{sen}}^2\text{Tr}\left(\boldsymbol{S}^H\dot{\boldsymbol{A}}_l^H\dot{\boldsymbol{A}}_l\boldsymbol{S}\right)\right]^{-1}$. The outage probability constraint for CRB, i.e., (\ref{eq14c}), is safely approximated in the following proposition (Proposition 3).

\noindent \textbf{Proposition 3.} \emph{The outage probability inequality (\ref{eq28}) can be safely approximated as}
\begin{equation}
        \label{eq29}
        \tilde{q}_l+\tilde{s}_l\ge2\sqrt{\ln(1/p_l)}\tilde{T}_l,\forall l\in\mathcal{L},
\end{equation}
\emph{where} $\tilde{T}_l=\tilde{v}_l\tilde{q}_l+\frac{1}{\sqrt{2}}\vert\tilde{r}_l\vert$, \emph{and} $\tilde{v}_l$ \emph{is the solution of the equation} $\left[1-1/\left(2\tilde{v}_l^2\right)\right]\tilde{v}_l=\sqrt{\ln\left(1/p_l\right)}$. 

\emph{Proof: }The proof is given in Appendix D.$\hfill\blacksquare$

The probability constraints (\ref{eq14b}) and (\ref{eq14c}) are safely approximated into the SOC constraints. We can rewritten $\text{Tr}\left(\boldsymbol{S}^H\dot{\boldsymbol{A}}_l^H\dot{\boldsymbol{A}}_l\boldsymbol{S}\right)$ as
\begin{subequations}
        \label{eq30}
        \begin{align}
                \text{Tr}\left(\boldsymbol{S}^H\dot{\boldsymbol{A}}_l^H\dot{\boldsymbol{A}}_l\boldsymbol{S}\right)&=\text{vec}(\boldsymbol{S})^H\left(\textbf{I}\otimes\dot{\boldsymbol{A}}_l^H\dot{\boldsymbol{A}}_l\right)\text{vec}(\boldsymbol{S})\nonumber\\
                &=\sum_{k=1}^{K}\boldsymbol{s}_k^H\dot{\boldsymbol{A}}_l^H\dot{\boldsymbol{A}}_l\boldsymbol{s}_k\nonumber\\
                &=\sum_{k=1}^{K}\text{Tr}\left(\boldsymbol{s}_k\boldsymbol{s}_k^H\dot{\boldsymbol{A}}_l^H\dot{\boldsymbol{A}}_l\right)\tag{30}
        \end{align}
\end{subequations}
Then, the beamforming problem can be rewritten as
\begin{subequations}
        \label{eq31}
        \begin{align}
                \min_{\boldsymbol{S},\boldsymbol{\theta},\boldsymbol{x},\boldsymbol{y}} \enspace & \Vert \boldsymbol{S} \Vert_F^2 \label{eq31a} \tag{31a}\\
                \text{s.t.}  \enspace& \left(\gamma_{\text{BU},k}^2+\gamma_{\text{BRU},k}^2M\right)\text{Tr}(\boldsymbol{\Psi}_k)+s_k\nonumber\\
                &\quad\geq2\sqrt{\ln(1/\rho_k)}(x_k+y_k),\forall k \in \mathcal{K},\label{eq31b} \tag{31b}\\
                &\sqrt{\frac{\left(\gamma_{\text{BU},k}^2+\gamma_{\text{BRU},k}^2M\right)}{2}}\bigg\Vert\left(\hat{\boldsymbol{h}}_{\text{BU},k}^H+\boldsymbol{\theta}^H\hat{\boldsymbol{H}}_{\text{BRU},k}\right)\nonumber\\
                &\cdot\boldsymbol{\Psi}_k\bigg\Vert\quad\leq x_k, \forall k \in \mathcal{K},\label{eq31c} \tag{31c}\\
                &v_k\left(\gamma_{\text{BU},k}^2+\gamma_{\text{BRU},k}^2M\right)\left\Vert\boldsymbol{\Psi}_k\right\Vert_F\leq y_k,\forall k \in \mathcal{K},\label{eq31d} \tag{31d}\\
                &\sum_{k=1}^{K}\text{Tr}\left\{\boldsymbol{s}_k\boldsymbol{s}_k^H\left(\dot{\boldsymbol{A}}_l^H\dot{\boldsymbol{A}}_l\right)\right\}\geq\nonumber\\
                &\Big[\varepsilon_l^2+\vert\hat{\alpha}_l\vert^2-2\sqrt{\ln(1/p_l)}\left(\tilde{v}_l\varepsilon_l^2+1/\sqrt{2}\vert\varepsilon_l\hat{\alpha}_l\vert\right)\nonumber\\
                &\cdot2c_l\sigma_{\text{sen}}^2\Big]^{-1},\forall l \in \mathcal{L},\label{eq31e} \tag{31e}\\
                &\theta_m\in\mathcal{F},\forall m \in\mathcal{M}.\label{eq31f} \tag{31f}
        \end{align}
\end{subequations}

\subsection{Transmit Beamforming}
The RIS beamforming matrix is fixed in this subsection. The transmit beamforming problem is formulated as a standard SDP problem, which can be solved by the SDR method. Specifically, the transmit beamforming problem can be written as
\begin{subequations}
        \label{eq32}
        \begin{align}
                \min_{\left\{\boldsymbol{\Gamma}_k\right\},\boldsymbol{x},\boldsymbol{y}} \enspace & \sum_{k=1}^{K}\text{Tr}\left\{\boldsymbol{\Gamma}_k\right\} \label{eq32a} \tag{32a}\\
                \text{s.t.}  \enspace& \text{(\ref{eq31b}), (\ref{eq31c}), (\ref{eq31d}), (\ref{eq31e}),}\label{eq32b} \tag{32b}\\
                &\boldsymbol{\Gamma}_k \succeq0,\forall k \in \mathcal{K},\label{eq32c} \tag{32c}\\
                &\text{rank}(\boldsymbol{\Gamma}_k)=1,\forall k \in \mathcal{K},\label{eq32d} \tag{32d}
        \end{align}
\end{subequations}
where $\boldsymbol{\Gamma}_k=\boldsymbol{s}_k\boldsymbol{s}_k^H$. The problem in (\ref{eq32a}) can be transformed to a SDP problem by omitting the (\ref{eq32d}) constraint. Then, the SDP problem can be solved by the CVX tool using the inter point method (IPM). By checking the Karush-Kuhn-Tucker (KKT) conditions of the problem in (\ref{eq32a}) without the rank one constraints, it is shown that the optimal beamforming matrices, i.e., $\boldsymbol{\Gamma}_k^*,\forall k\in\mathcal{K}$, satisfying $\text{rank}(\boldsymbol{\Gamma}_k^*)=1$ can always be obtained. This means we can always find the optimal BS beamforming matrix through the SDR method.

\subsection{RIS Beamforming}
We solve the RIS beamforming problem in this subsection. The transmit beamforming matrix is kept fixed. The RIS beamforming problem is a feasibility-check problem and a non-convex optimization problem due to the integer constraint (\ref{eq31f}) for the discrete RIS phase shifts. We propose two schemes to efficiently obtain the RIS beamforming vector, which are as follows.

\subsubsection{Scheme 1}
To accelerate the convergence of the AO algorithm, we transform the RIS beamforming problem as
\begin{subequations}
    \label{eq33}
    \begin{align}
            \max_{\boldsymbol{\theta},\boldsymbol{x},\boldsymbol{\bar{\alpha}}} \enspace & \sum_{k=1}^{K}\bar{\alpha}_k \label{eq33a} \tag{33a}\\
            \text{s.t.}  \enspace& \left(\gamma_{\text{BU},k}^2+\gamma_{\text{BRU},k}^2M\right)\text{Tr}(\boldsymbol{\Psi}_k)+\bar{s}_k\nonumber\\
            &\quad\geq2\sqrt{\ln(1/\rho_k)}(x_k+\bar{y}_k),\forall k \in \mathcal{K},\label{eq33b} \tag{33b}\\
            &\bar{\alpha}_k\geq0,\forall k \in \mathcal{K}\label{eq33c} \tag{33c},\\
            &\text{(\ref{eq31c}), (\ref{eq31f}),}\label{eq33d} \tag{33d}
    \end{align}
\end{subequations}
where $\boldsymbol{\bar{\alpha}}\triangleq[\bar{\alpha}_1,\cdots,\bar{\alpha}_K]^T$ is a slack vector variable, $\bar{s}_k = s_k-\bar{\alpha}_k$, and $\bar{y}_k=v_k\left(\gamma_{\text{BU},k}^2+\gamma_{\text{BRU},k}^2M\right)\left\Vert\boldsymbol{\Psi}_k\right\Vert_F$ according to (\ref{eq31d}). 

The problem in (\ref{eq33a}) is still non-convex due to the non-homogeneous quadratic constraints (\ref{eq33b}) and (\ref{eq31c}). We reformulate $\bar{s}_k$ in (\ref{eq33b}) as
\begin{subequations}
    \label{eq34}
    \begin{align}
            \bar{s}_k&=\bar{\boldsymbol{\theta}}^H\bar{\boldsymbol{G}}_k\bar{\boldsymbol{\theta}}+\hat{\boldsymbol{h}}_{\text{BU},k}^H\boldsymbol{\Psi}_k\hat{\boldsymbol{h}}_{\text{BU},k}-\sigma_{\text{sen}}^2-\bar{\alpha}_k\nonumber\\
            &=\text{Tr}(\bar{\boldsymbol{G}}_k\bar{\boldsymbol{\Theta}})+\hat{\boldsymbol{h}}_{\text{BU},k}^H\boldsymbol{\Psi}_k\hat{\boldsymbol{h}}_{\text{BU},k}-\sigma_{\text{sen}}^2-\bar{\alpha}_k,\tag{34}
    \end{align}
\end{subequations}
where
\begin{equation}
        \label{eq35}
        \bar{\boldsymbol{\theta}}=\begin{bmatrix}
                \boldsymbol{\theta} \\
                t
        \end{bmatrix},
        \bar{\boldsymbol{G}}_{k} = \begin{bmatrix}
                \hat{\boldsymbol{H}}_{\text{BRU},k}\boldsymbol{\Psi}_k\hat{\boldsymbol{H}}_{\text{BRU},k}^H, &\!\!\!\!\! \hat{\boldsymbol{H}}_{\text{BRU},k}\boldsymbol{\Psi}_k\hat{\boldsymbol{h}}_{\text{BU},k}\\
                \hat{\boldsymbol{h}}_{\text{BU},k}^H\boldsymbol{\Psi}_k\hat{\boldsymbol{H}}_{\text{BRU},k}^H, &\!\!\!\!\! 0
        \end{bmatrix},
\end{equation}
$\bar{\boldsymbol{\Theta}}=\bar{\boldsymbol{\theta}}\bar{\boldsymbol{\theta}}^H$, and $t$ is an auxiliary variable. Then, (\ref{eq31c}) is rewritten as
\begin{subequations}
        \label{eq36}
        \begin{align}
                &\text{Tr}\left(\tilde{\boldsymbol{G}}_k\bar{\boldsymbol{\Theta}}\right) + \hat{\boldsymbol{h}}_{\text{BU},k}^H\boldsymbol{\Psi}_k\boldsymbol{\Psi}_k^H\hat{\boldsymbol{h}}_{\text{BU},k}
                \nonumber\\
                &\quad\leq 2x_k^2/\left(\gamma_{\text{BU},k}^2+\gamma_{\text{BRU},k}^2M\right), \forall k \in \mathcal{K}\tag{36}
        \end{align}
\end{subequations}
where 
\begin{equation}
	\label{eq37}
	\tilde{\boldsymbol{G}}_{k} = \begin{bmatrix}
		\hat{\boldsymbol{H}}_{\text{BRU},k}\boldsymbol{\Psi}_k\boldsymbol{\Psi}_k^H\hat{\boldsymbol{H}}_{\text{BRU},k}^H & \hat{\boldsymbol{H}}_{\text{BRU},k}\boldsymbol{\Psi}_k\boldsymbol{\Psi}_k^H\hat{\boldsymbol{h}}_{\text{BU},k}\\
		\hat{\boldsymbol{h}}_{\text{BU},k}^H\boldsymbol{\Psi}_k\boldsymbol{\Psi}_k^H\hat{\boldsymbol{H}}_{\text{BRU},k}^H & 0
	\end{bmatrix}.
\end{equation}
The RIS beamforming problem is formulated as a SDP problem, which is expressed as
\begin{subequations}
        \label{eq38}
        \begin{align}
                \max_{\bar{\boldsymbol{\Theta}},\boldsymbol{x},\boldsymbol{\bar{\alpha}},t} \enspace & \sum_{k=1}^{K}\bar{\alpha}_k \label{eq38a} \tag{38a}\\
                \text{s.t.}  \enspace& \text{(\ref{eq33b}), (\ref{eq33c}), (\ref{eq36}),}\label{eq38b} \tag{38b}\\
                &\bar{\boldsymbol{\Theta}}_{m,m}=1,\forall m \in\mathcal{M},\label{eq38c} \tag{38c}\\
                &\bar{\boldsymbol{\Theta}}\succeq0.\label{eq38d} \tag{38d}
        \end{align}
\end{subequations}
Since the SDP problem is not guaranteed to obtain a beamforming matrix with rank on, a Gaussian randomization algorithm is proposed to ensure a rank one solution. We present the Gaussian randomization algorithm in \textbf{Algorithm 1}.

Note that we relax the constraint (\ref{eq31f}) as a unit-modulus one. The obtained continuous RIS beamforming vector are mapped to a discrete one, which is expressed as
\begin{equation}
        \label{eq39}
        \theta_m^*=\text{arg }\min_{\theta\in\mathcal{F}}\vert\theta_m-\theta\vert,\forall m \in\mathcal{M},
\end{equation}
where $\theta_m^*$ is the desired discrete phase shift and $\theta_m$ is the continuous phase shift obtained by solving the problem in (\ref{eq38a}).
\begin{algorithm}[t]
	\caption{Gaussian Randomization Method.}\label{alg:alg1}
	\small
	\begin{algorithmic}
		\STATE 
		\STATE 1. $ \textbf{Inputs: }$$\boldsymbol{\bar{\Theta}}$,$\hat{\boldsymbol{h}}_{\text{BU},k}$, $\hat{\boldsymbol{H}}_{\text{BRU},k}$,$\hat{\alpha}_l$, $\gamma_{\text{BU},k}^2$, $\gamma_{\text{BRU},k}^2$, $\varepsilon_l^2$, $r_k$, $\rho_k$, $p_l$, $\phi_l$, $\sigma_{\text{com},k}^2$, $\sigma_{\text{sen}}^2$, $G_{\text{max}}$. 
		\STATE 2. $ \textbf{Outputs: }$Beamforming matrix $\boldsymbol{\Theta}^*$ for (\ref{eq33a}).
		\STATE 3. Eignvalue decompose $\boldsymbol{\bar{\Theta}}$, i.e., $\boldsymbol{\bar{\Theta}} = \boldsymbol{\Sigma}\boldsymbol{V}$, where $\boldsymbol{\Sigma}$ is the eignvalue matrix and $\boldsymbol{V}$ is the eigenvector matrix.
		\FOR{$i=1$ to $G_{\text{max}}$}
		\STATE 4. Randomly generate $\tilde{\boldsymbol{x}}_i$, which is assume to follow the standard CSCG distribution. And let $\tilde{\boldsymbol{\theta}}_i = \boldsymbol{V}\boldsymbol{\Sigma}^{1/2}\tilde{\boldsymbol{x}}_i $.
		\IF{$\tilde{\boldsymbol{\theta}}_i$ makes the communication rate bigger}
		\STATE 5. Let $\boldsymbol{\Theta}^* =\text{diag}\left(\tilde{\boldsymbol{\theta}}_i\right) $.
		\ENDIF
		\ENDFOR
	\end{algorithmic}
	\label{alg1}
\end{algorithm}

\subsubsection{Scheme 2}
In this scheme, the RIS beamforming problem is transformed into a SP problem in order to accelerate the convergence of the AO algorithm. The SP problem is expressed as
\begin{subequations}
        \label{eq40}
        \begin{align}
                \max_{\boldsymbol{\theta}} \enspace & \sum_{k=1}^{K}\bigg[s_k-\sqrt{2\ln(1/\rho_k)(\gamma_{\text{BU},k}^2+\gamma_{\text{BRU},k}^2M)}\nonumber\\
                &\quad\enspace\left\Vert \left(\hat{\boldsymbol{h}}_{\text{BU},k}^H+\boldsymbol{\theta}^H\hat{\boldsymbol{H}}_{\text{BRU},k}\right)\boldsymbol{\Psi}_k\right\Vert\bigg]+\lambda\Vert \boldsymbol{\theta}\Vert^2 \label{eq40a} \tag{40a}\\
                \text{s.t.}&\enspace \theta_m \in \tilde{\mathcal{F}},\forall m \in \mathcal{M}, \label{eq40b} \tag{40b}
        \end{align}
\end{subequations}
where $\tilde{\mathcal{F}}$ is the convex hull of $\mathcal{F}$, i.e., $\tilde{\mathcal{F}}$ is a regular polygon with $\mathcal{F}$ as its vertices. Note that the phase shifts for RIS elements will be located at the vertices of the regular polygon when the penalty factor $\lambda$ is sufficiently large. The objective function in (\ref{eq40a}) is formulated from (\ref{eq31b}) and (\ref{eq31c}), where the constant terms, such as $\text{Tr}(\boldsymbol{\Psi}_k)$ and $y_k$, are omitted. The basic idea of (\ref{eq40a}) is to minimize the outage probability of all communication users. Then, we use GEMM method to solve the SP problem. The GEMM method is an efficient algorithm based on gradient ascent to solve the smooth but non-convex problem \cite{ref46}. 

Let $f_\lambda(\boldsymbol{\theta})$ denotes the objective function in (\ref{eq40a}) except for the term $\lambda\Vert \boldsymbol{\theta}\Vert^2$. In order to apply GEMM method, the majorant function of $f_\lambda(\boldsymbol{\theta})$ should be firstly found, i.e., we need to find a majorant function $F_\lambda(\boldsymbol{\theta}\vert\tilde{\boldsymbol{\theta}})$ to approximate $f_\lambda(\boldsymbol{\theta})$ at point $\tilde{\boldsymbol{\theta}}$. The majorant function $F_\lambda(\boldsymbol{\theta}\vert\tilde{\boldsymbol{\theta}})$ needs to satisfy the following three condition
\begin{itemize}
        \item$F_\lambda(\boldsymbol{\theta}\vert\tilde{\boldsymbol{\theta}})\leq f_\lambda(\boldsymbol{\theta}),\forall \boldsymbol{\theta},\tilde{\boldsymbol{\theta}}\in\tilde{\mathcal{F}}$;
        \item $F_\lambda(\tilde{\boldsymbol{\theta}}\vert\tilde{\boldsymbol{\theta}})=f_\lambda(\tilde{\boldsymbol{\theta}}),\forall \tilde{\boldsymbol{\theta}}\in\tilde{\mathcal{F}}$;
        \item$\nabla_{\boldsymbol{\theta}}F_\lambda(\boldsymbol{\theta}\vert\tilde{\boldsymbol{\theta}})=\nabla_{\boldsymbol{\theta}}f_\lambda(\boldsymbol{\theta}),\forall \boldsymbol{\theta},\tilde{\boldsymbol{\theta}}\in\tilde{\mathcal{F}}$.
\end{itemize}
It is not difficult to find the majorant function of $f_\lambda(\boldsymbol{\theta})$ that satisfies all the above conditions. We have
\begin{equation}
        \label{eq41}
        F_\lambda(\boldsymbol{\theta}\vert\tilde{\boldsymbol{\theta}}) \triangleq f_\lambda(\boldsymbol{\theta})+\lambda\left(\Vert\tilde{\boldsymbol{\theta}}\Vert^2 + 2\left<\tilde{\boldsymbol{\theta}},\boldsymbol{\theta}-\tilde{\boldsymbol{\theta}}\right>\right).
\end{equation}

Further, the SP problem (\ref{eq40a}) can be solved by GEMM method in an iterative manner, which can be expressed as
\begin{subequations}
        \label{eq42}
        \begin{align}
                \boldsymbol{\theta}^{(i+1)} = \arg \min_{\boldsymbol{\theta}} \quad & F_\lambda(\boldsymbol{\theta} \vert \boldsymbol{\theta}^{(i)}) \label{eq42a} \tag{42a}\\
                \text{s.t.} \quad & \text{(\ref{eq40b})},  \label{eq42b} \tag{42b}
        \end{align}
\end{subequations}
where $i$ denotes the iterative number. We use the APG method to solve (\ref{eq42}), since the APG method has better convergence performance than the projected gradient (PG). 

Specifically, the APG method approaches the optimal solution by iteratively updating
\begin{equation}
        \label{eq43}
        \boldsymbol{\theta}^{(i+1)} =  \Pi_{\tilde{\mathcal{F}}}\left( \boldsymbol{z}^{(i)} - 
        \frac{1}{\beta^{(i)}}\nabla_{\boldsymbol{\theta}}F_\lambda(\boldsymbol{z}^{(i)} \vert \boldsymbol{\theta}^{(i)}  \right), 
\end{equation}
where $\Pi_{\tilde{\mathcal{F}}}$ is the projection of $\boldsymbol{\theta}$ onto $\tilde{\mathcal{F}}$, $\beta^{(i)}$ denotes the step length, and $\boldsymbol{z}^{(i)}$ denotes the extrapolated point of the majorant function at the $i$-th iteration. It should be noted that a large number of iterations are needed to solve (\ref{eq42}). Considering the computational complexity, we only utilize one step APG method to update (\ref{eq43}). Moreover, $\boldsymbol{z}^{(i)}$ is computed by
\begin{equation}
        \label{eq44}
        \boldsymbol{z}^{(i)} = \boldsymbol{\theta}^{(i)} + \alpha_i\left( \boldsymbol{\theta}^{(i)} - \boldsymbol{\theta}^{(i-1)} \right), 
\end{equation}
where
\begin{equation}
        \label{eq45}
        \alpha_i \triangleq \frac{\xi^{(i-1)}-1}{\xi^{(i)}}, 
\end{equation}
and
\begin{equation}
        \label{eq46}
        \xi^{(i)} \triangleq \frac{1+\sqrt{1+4\left(\xi^{(i-1)}\right)^2}}{2}
\end{equation}
with $\boldsymbol{\theta}^{(-1)} = \boldsymbol{\theta}^{(0)}$ and $\xi^{(-1)}=0$. The projection $\Pi_{\tilde{\mathcal{F}}}$ operation is defined as
\begin{subequations}
        \label{eq47}
        \begin{align}
        \Pi_{\tilde{\mathcal{F}}} (\boldsymbol{\theta}) \triangleq e^{j\frac{2\pi n}{d}} \bigg(& \left[ \text{Re}\left\{ \hat{\boldsymbol{\theta}} \right\} \right]_0^{\cos(\pi/d)} \nonumber \\
        &+ 
        j\left[ \text{Im}\left\{ \hat{\boldsymbol{\theta}} \right\} \right]_{-\sin(\pi/d)}^{\sin(\pi/d)} \bigg),\tag{47}
   	 	\end{align}
\end{subequations}
where $n=\lfloor \frac{\angle \boldsymbol{\theta}+\pi/d}{2\pi/d} \rfloor$, $\hat{\boldsymbol{\theta}}=\boldsymbol{\theta}e^{-j\frac{2\pi n}{d}}$. The step length $\beta^{(i)}$ needs to satisfy the ascent condition, which is written as
\begin{subequations}
        \label{eq48}
        \begin{align}
                F_\lambda(\boldsymbol{\theta}^{(i+1)} \vert \boldsymbol{\theta}^{(i)}) &\geq F_\lambda(\boldsymbol{z}^{(i)} \vert \boldsymbol{\theta}^{(i)}) + \frac{\beta^{(i)}}{2} \left\Vert \boldsymbol{\theta}^{(i+1)} - \boldsymbol{z}^{(i)} \right\Vert^2 \nonumber \\
                &+ \left<\nabla_{\boldsymbol{\theta}}F_\lambda(\boldsymbol{z}^{(i)} \vert \boldsymbol{\theta}^{(i)}), \boldsymbol{\theta}^{(i+1)} - \boldsymbol{z}^{(i)}\right>, \tag{48}
        \end{align}
\end{subequations}
where the gradient $\nabla_{\boldsymbol{\theta}}F_\lambda$ is derived as
\begin{subequations}
        \label{eq49}
        \begin{align}
                &\nabla_{\boldsymbol{\theta}}F_\lambda(\boldsymbol{z}^{(i)} \vert \boldsymbol{\theta}^{(i)})=\nabla_{\boldsymbol{\theta}}f_\lambda(\boldsymbol{\theta})\bigg|_{\boldsymbol{\theta} = \boldsymbol{z}^{(i)}}+2\lambda\boldsymbol{\theta}^{(i)}\nonumber\\
                &=\sum_{k=1}^{K}\Bigg[2\left(\hat{\boldsymbol{H}}_{\text{BRU},k}\boldsymbol{\Psi}_k\hat{\boldsymbol{h}}_{\text{BU},k}+\hat{\boldsymbol{H}}_{\text{BRU},k}\boldsymbol{\Psi}_k\hat{\boldsymbol{H}}_{\text{BRU},k}^H\boldsymbol{z}^{(i)}\right)\nonumber\\
                &\quad-\sqrt{2\ln(1/\rho_k)(\gamma_{\text{BU},k}^2+\gamma_{\text{BRU},k}^2M)}\nonumber\\
                &\quad\cdot\frac{\hat{\boldsymbol{H}}_{\text{BRU},k}\boldsymbol{\Psi}_k^2\hat{\boldsymbol{h}}_{\text{BU},k}+\hat{\boldsymbol{H}}_{\text{BRU},k}\boldsymbol{\Psi}_k^2\hat{\boldsymbol{H}}_{\text{BRU},k}^H\boldsymbol{z}^{(i)}}{\left\Vert \left(\hat{\boldsymbol{h}}_{\text{BU},k}^H+(\boldsymbol{z}^{(i)})^H\hat{\boldsymbol{H}}_{\text{BRU},k}\right)\boldsymbol{\Psi}_k\right\Vert}\Bigg]+2\lambda\boldsymbol{\theta}^{(i)}.\tag{49}
        \end{align}
\end{subequations}

The proposed algorithm for solving the RIS beamforming problem (\ref{eq40}) based on GEMM method is presented in \textbf{Algorithm 2}. The proposed AO algorithm for solving the robust beamforming problem (\ref{eq14}) is presented in \textbf{Algorithm 3}.
\begin{algorithm}[t]
        \caption{SP-Based GEMM Method to Solve (\ref{eq40a}).}\label{alg:alg2}
        \small
        \begin{algorithmic}
                \STATE 
                \STATE 1. $ \textbf{Inputs: }$$\hat{\boldsymbol{h}}_{\text{BU},k}$, $\hat{\boldsymbol{H}}_{\text{BRU},k}$, $\boldsymbol{S}$, $\boldsymbol{\Psi}_k$, $\boldsymbol{\theta}^{(0)}$, $\sigma_{\text{com},k}^2$, $\rho_k$, $\gamma_{\text{BU},k}^2$, $\gamma_{\text{BRU},k}^2$, $\lambda$, $i_{\text{max}}$. 
                \STATE 2. $ \textbf{Outputs: }$RIS beamforming vector $\boldsymbol{\theta}^*$ for (\ref{eq40a}).
                \STATE 3. $ \textbf{Initialization: }$ Set $\boldsymbol{\theta}^{(-1)}=\boldsymbol{\theta}^{(0)}$ and $\xi^{(-1)}=0$.
                \WHILE {$i \leq i_{\text{max}}$, and (\ref{eq48}) is satisfied}
                \STATE 4. Compute $\xi^{(i)}$ by (\ref{eq46}).
                \STATE 5. Compute $\alpha^{(i)}$ by (\ref{eq45}).
                \STATE 6. Compute $\boldsymbol{z}^{(i)}$ by (\ref{eq44}).
                \STATE 7. Obtain the gradient $\nabla_{\boldsymbol{\theta}}F_\lambda$ using (\ref{eq49}).
                \STATE 8. Choose the step length $\beta^{(i)}$ by (\ref{eq48}).
                \STATE 9. Obtain $\boldsymbol{\theta}^{(i+1)}$ by (\ref{eq43}) and (\ref{eq47}).
                \STATE 10. $i=i+1$.
                \ENDWHILE
                \STATE 11. $\boldsymbol{\theta}^* = \boldsymbol{\theta}^{(i)}$.
        \end{algorithmic}
        \label{alg2}
\end{algorithm}
\begin{algorithm}[t]
        \caption{Alternating Optimization (AO) Algorithm to Minimize the Transmit Power.}\label{alg:alg3}
        \small
        \begin{algorithmic}
                \STATE 
                \STATE 1. $ \textbf{Inputs: }$$\hat{\boldsymbol{h}}_{\text{BU},k}$, $\hat{\boldsymbol{H}}_{\text{BRU},k}$,$\hat{\alpha}_l$, $\gamma_{\text{BU},k}^2$, $\gamma_{\text{BRU},k}^2$, $\varepsilon_l^2$, $r_k$, $\rho_k$, $p_l$, $\phi_l$, $\sigma_{\text{com},k}^2$, $\sigma_{\text{sen}}^2$, $i_{\text{max}}$, $\epsilon$. 
                \STATE 2. $ \textbf{Outputs: }$Transmit beamforming matrix $\boldsymbol{S}^*$ and RIS beamforming vector $\boldsymbol{\theta}^*$.
                \STATE 3. $ \textbf{Initialization: }$ Randomly initialize $\boldsymbol{S}^{\left(1\right)} \text{,} \boldsymbol{\theta}^{\left(1\right)}$, compute the objective function in (\ref{eq14a}) as $\delta^{\left(1\right)}$, set $ \delta^{\left(0\right)}=+\infty,i=0 $.
                \STATE 4. \textbf{Approximation: }Approximate (\ref{eq14b}) by (\ref{eq24}), (\ref{eq25}), and (\ref{eq26}). Approximate (\ref{eq14c}) by (\ref{eq29}).
                \WHILE {$i \leq i_{\text{max}}$, and $\delta^{\left(i-1\right)}-\delta^{\left(i\right)} \geq \epsilon$}
                \STATE 5. Fix $\boldsymbol{\theta}^{\left(i\right)}$, solve the convex problem (\ref{eq32a}).
                \IF {Constraint (\ref{eq32d}) is satisfied}
                \STATE 6. Obtain $ \boldsymbol{S}^{\left(i+1\right)} $ directly.
                \ELSE
                \STATE 7. Perform Gaussian randomization to obtain $ \boldsymbol{S}^{\left(i+1\right)} $.
                \ENDIF
                \STATE 8. Fix $\boldsymbol{S}^{\left(i+1\right)}$, solve the convex problem (\ref{eq38a}) in scheme 1, or solve the problem (\ref{eq40a}) in scheme 2 to obtain $\boldsymbol{\theta}^{(i+1)}$.
                \IF {scheme 1 is choosen}
                \STATE 9. Perform the mapping operation (\ref{eq39}) to obtain $\boldsymbol{\theta}^{(i+1)}$.
                \ENDIF
                \STATE 10. $i=i+1$.
                \ENDWHILE
                \STATE 11. $\boldsymbol{S}^* = \boldsymbol{S}^{\left(i\right)}$, $\boldsymbol{\theta}^* = \boldsymbol{\theta}^{\left(i\right)}$.
        \end{algorithmic}
        \label{alg3}
\end{algorithm}
        
\section{Complexity and Convergence Analysis}
In this section, we analyze the computational complexity of the SDP methods for solving the transmit beamforming problem and the RIS beamforming problem in scheme 1. Moreover, the SP-based GEMM method in scheme 2 is also analyzed. The overall complexity of the proposed AO algorithm will be verified by simulation results in the next section since the number of the required total iteration is hard to predict. We use the flops number to denote the computational complexity. A flop is defined as one real addition or one real multiplication \cite{ref5}.

\subsection{Complexity Analysis for SDP Problems}
The SDP problems with $n$ constraints and the dimension of variables in each constraint equals to $m\times m$ can be solved with a worst-case complexity of $\mathcal{O}\left(\log\left(1/\eta\right)\max\left\{m,n\right\}^4m^{0.5}\right)$, where $\eta$ is the accuracy of solution \cite{ref48}. Therefore, the SDP problems (\ref{eq32a}) and (\ref{eq38a}) can be solved with a worst-case complexity of $\mathcal{O}_{\text{SDP}}^{\text{Transmit}}=\mathcal{O}\left(\log\left(1/\eta\right)\max\left\{N,4K+L\right\}^4N^{0.5}\right)$ and $\mathcal{O}_{\text{SDP}}^{\text{RIS}}=\mathcal{O}\left(\log\left(1/\eta\right)\max\left\{M+1,3K+2\right\}^4(M+1)^{0.5}\right)$, respectively.

\subsection{Complexity Analysis for the SP Problem}
The complexity of the GEMM method mainly includes line 7, line 8, and line 9 in \textbf{Algorithm 1}. The number of flops for calculating line 7, line 8, and line 9 are $48MN^2+16M^2N+24MN+16M^2+8N^2+2(K+6)M+10N$, $24MNK+16N^2K+20MK+18M+22N$, and $6M$, respectively. Therefore, the complexity of the GEMM method is $\mathcal{O}_{\text{GEMM}}^{\text{RIS}}=\mathcal{O}\left(MN^2+M^2N+MNK+N^2K\right)$.

\subsection{Convergence Analysis for the AO-SDR Algorithm}
In this subsection, we analyze the convergence performance of the proposed AO-SDR algorithm, i.e., the AO based on the SDR method and the RIS beamforming scheme 1. Firstly, the transmit power obtained by the proposed AO-SDR algorithm has a lower bound due to user rate constraints. We assume the lower bound of the transmit power to be $P_{\text{min}}$. Next, we prove that with the iterations of the AO algorithm, the transmit power decreases. Denoting the transmit power at the $i$-th AO iteration by $P\left(\boldsymbol{S}[i],\boldsymbol{\theta}[i]\right)$. We have
\begin{equation}
	\label{eq50}
	\begin{aligned}								P_{\text{min}}&\overset{\text{(a)}}{\leq}P\left(\boldsymbol{S}[i+1],\boldsymbol{\theta}[i+1]\right)\\
		&\overset{\text{(b)}}{\leq}P\left(\boldsymbol{S}[i],\boldsymbol{\theta}[i+1]\right)\overset{\text{(c)}}{\leq}P\left(\boldsymbol{S}[i],\boldsymbol{\theta}[i]\right),
	\end{aligned}
\end{equation}
where $(a)$ holds since the transmit power has a lower bound, $(b)$ holds since $\boldsymbol{S}[i+1]$ is the optimal BS beamforming matrix of the $i+1$-th AO iteration; $(c)$ holds because the transmit power only depends on $\boldsymbol{S}$.

\subsection{Convergence Analysis for the AO-GEMM Algorithm}
In this subsection, we analyze the convergence performance of the proposed AO-GEMM algorithm, i.e., the AO based on the SDR method and the RIS beamforming scheme 2. Since the analysis is similar with that of the last subsection, we omit the convergence analysis of the AO-GEMM algorithm.

\subsection{Discussion}
The complexity of the SDR method for solving the transmit beamforming scales approximately with the fourth power of the number of BS antennas, communication users, and sensing targets. When the numbers of BS antennas, users, and targets is large, considerable computational resources are required to execute the SDR method. In cases where computational resources are insufficient, another alternative suboptimal algorithm is needed to obtain the transmit beamforming matrix. Moreover, the complexity of the SDR method for solving the RIS beamforming scales approximately with the fourth power of the number of RIS elements and communication users. In cases where computational resources are insufficient, we recommend using the proposed GEMM algorithm, i.e., RIS beamforming scheme 2, to solve the RIS beamforming problem.

\section{Simulation Results}
All communication channel model, i.e., $\boldsymbol{h}_{\text{BU},k},\boldsymbol{h}_{\text{RU},k},\forall k \in\mathcal{K}$, and $\boldsymbol{H}_{\text{BR}}$ are assumed to include the large-scale path loss and the small-scale fading. The large-scale path loss is expressed as $\text{Pathloss}\triangleq-30-10p\log_{10}(d)$dB with $p$ denoting the path loss factor and $d$ denoting distance. The path loss factor for $\boldsymbol{h}_{\text{BU},k},\boldsymbol{h}_{\text{RU},k}$, and $\boldsymbol{H}_{\text{BR}}$ are 4, 2, and 2.2, respectively. The small-scale fading is assumed to follow Rician fading channel model. For example, the BS-RIS channel $\boldsymbol{H}_{\text{BR}}$ is written as
\begin{equation} 
        \label{eq99}
        \boldsymbol{H}_{\text{BR}} = \sqrt{\frac{\beta_{\text{BR}}}{1+\beta_{\text{BR}}}}\boldsymbol{H}_{\text{BR}}^{\text{LOS}}+\sqrt{\frac{1}{1+\beta_{\text{BR}}}}\boldsymbol{H}_{\text{BR}}^{\text{NLOS}}, 
\end{equation}
where $\boldsymbol{H}_{\text{BR}}^{\text{LOS}}$ is the deterministic line-of-sight (LoS) component with $\beta_{\text{BR}}$ denoting the Rician factor, and $\boldsymbol{H}_{\text{BR}}^{\text{NLOS}}$ is the random Rayleigh fading component, which follows the CSCG distribution. The Rician factor for $\boldsymbol{h}_{\text{BU},k},\boldsymbol{h}_{\text{RU},k}$, and $\boldsymbol{H}_{\text{BR}}$ are 0, 0, and $\infty$, respectively. The radar channel coefficient $\alpha_l$ is assumed to follow the CSCG distribution due to the fluctuation of target, which is expressed as $\alpha_l \sim \mathcal{CN}\left(0,\sqrt{\frac{\lambda^2\kappa}{64\pi^3\left(d_l^{\text{RT}}\right)^4}}\right)$, where $\lambda$ is the wavelength, $\kappa$ is the radar cross-section (RCS) of the target, and $d_l^{\text{RT}}$ is the distance between the RIS and the $l$-th target.

\begin{figure}[!t]
	\centering
	\includegraphics[width=0.45\textwidth]{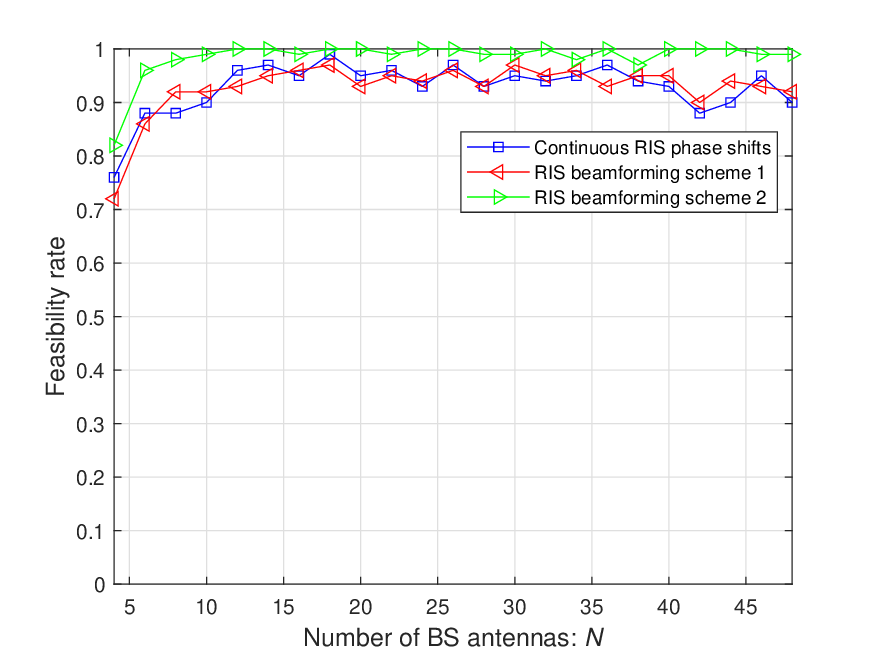}
	\caption{Feasibility rate of the proposed AO algorithm versus the number of BS antennas when $\gamma_{\text{BU},k}, \gamma_{\text{BRU},k}$, and $\varepsilon_l$ both equal 0.01.}
	\label{fig_9}
\end{figure}

\begin{figure}[!t]
	\centering
	\includegraphics[width=0.45\textwidth]{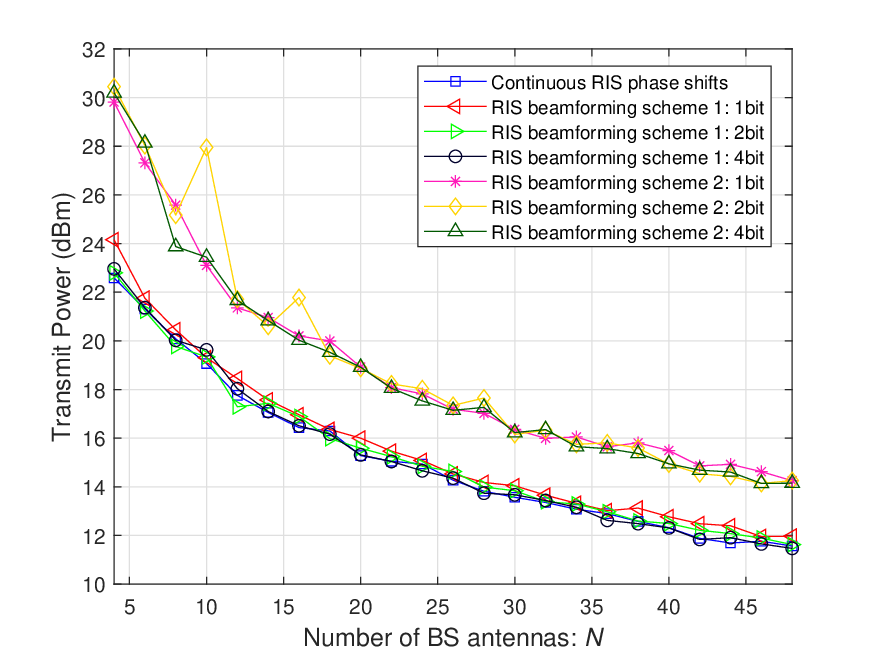}
	\caption{Transmit power versus the number of BS antennas when $\gamma_{\text{BU},k}, \gamma_{\text{BRU},k}$, and $\varepsilon_l$ both equal 0.01.}
	\label{fig_10}
\end{figure}

\begin{figure}[!t]
	\centering
	\includegraphics[width=0.45\textwidth]{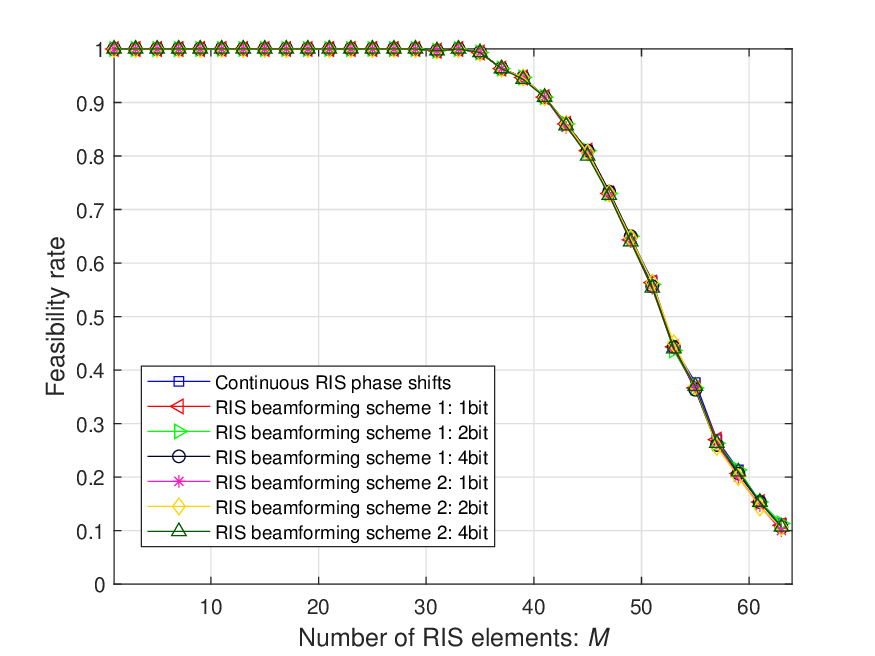}
	\caption{Feasibility rate of the proposed AO algorithm versus the number of RIS elements when $\gamma_{\text{BU},k}, \gamma_{\text{BRU},k}$, and $\varepsilon_l$ both equal 0.01.}
	\label{fig_2}
\end{figure}

\begin{figure}[!t]
	\centering
	\includegraphics[width=0.45\textwidth]{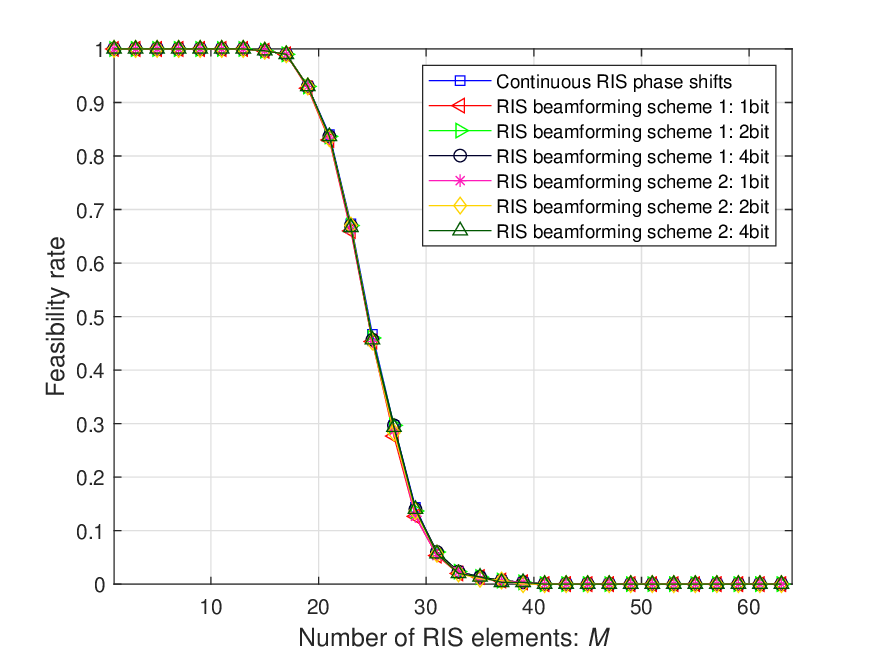}
	\caption{Feasibility rate of the proposed AO algorithm versus the number of RIS elements when $\gamma_{\text{BU},k}, \gamma_{\text{BRU},k}$, and $\varepsilon_l$ both equal 0.02.}
	\label{fig_3}
\end{figure}

\begin{figure}[!t]
	\centering
	\includegraphics[width=0.45\textwidth]{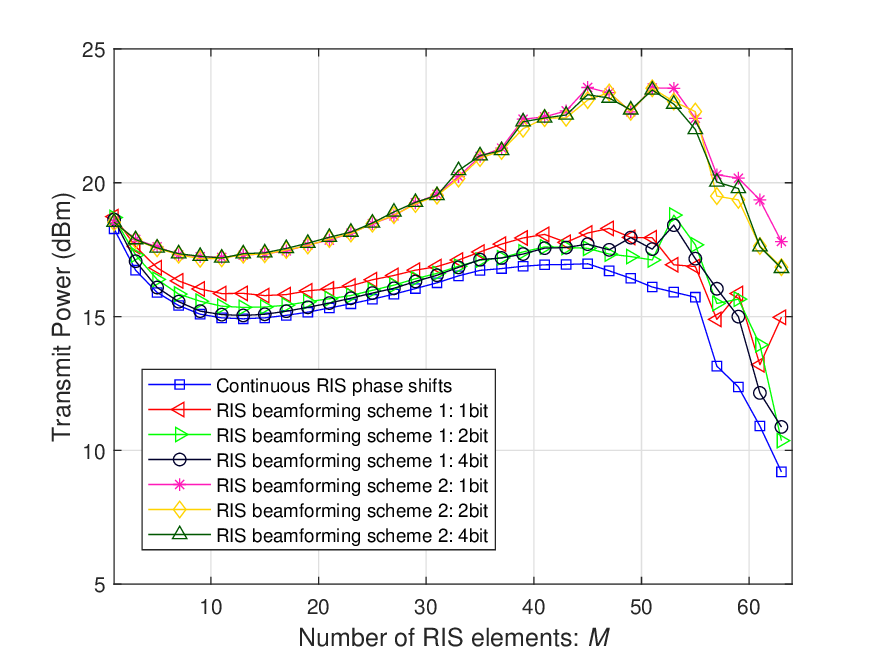}
	\caption{Transmit power versus the number of RIS elements when $\gamma_{\text{BU},k}, \gamma_{\text{BRU},k}$, and $\varepsilon_l$ both equal 0.01.}
	\label{fig_4}
\end{figure}

\begin{figure}[!t]
	\centering
	\includegraphics[width=0.45\textwidth]{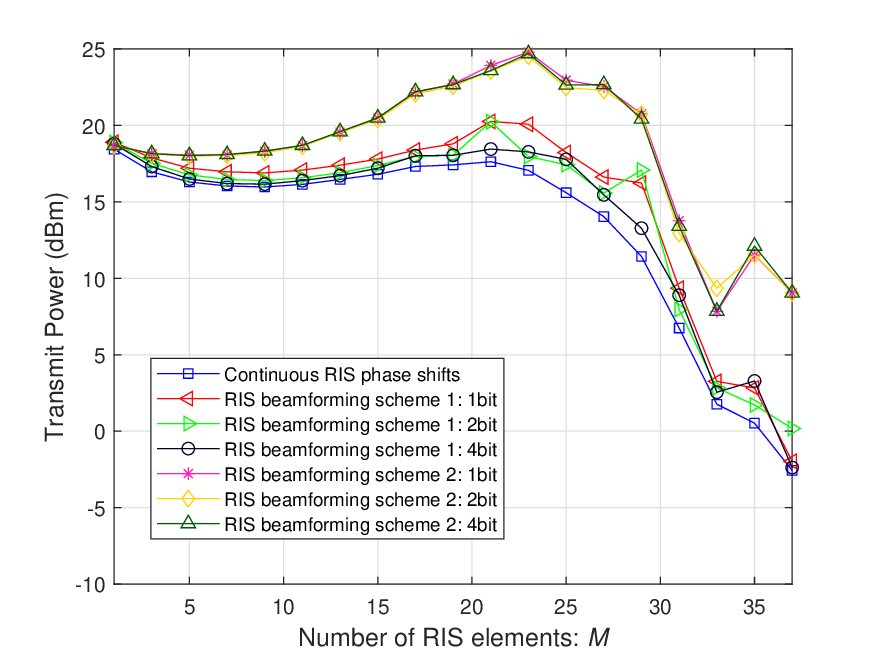}
	\caption{Transmit power versus the number of RIS elements when $\gamma_{\text{BU},k}, \gamma_{\text{BRU},k}$, and $\varepsilon_l$ both equal 0.02.}
	\label{fig_5}
\end{figure}

We assume that the BS is located at(0m, 0m); the RIS is located at (50m,10m); the $K$ users and $L$ targets are uniformly distributed within a circle centered at (70m,0m) with a radius of 5 meters and within a circle centered at the base station with a radius of 70 meters. Unless otherwise specified, the number of ISAC BS antennas is 16, the number of targets is 2, the number of users is 2, the communication rate threshold $R_0$ equals 2 bits/s/Hz, the CRB threshold equals 0.01, the probability $\rho_k$ and $p_l$ both equal 0.05, and the noise power is -110dBm. Moreover, from Fig. \ref{fig_9} to Fig. \ref{fig_5}, the transmit power obtained based on the AO-SDR algorithm and continuous RIS phase shifts is taken as the lower bound, labelled as ``Continuous RIS phase shifts'' because the AO-SDR algorithm has been shown in \cite{ref11} to achieve a near-optimal solution.

Figure \ref{fig_9} illustrates that the feasibility of all schemes is always high. This is because the BS active beamforming provides higher DoF to compensate for CSI errors, especially when the number of BS antennas is large.

Figure \ref{fig_10} shows that the required transmit power decreases as the number of BS antennas increases. This is because the beamforming gain provided by the BS antennas increases as the number of antennas increases, exceeding the power loss caused by compensating for CSI errors.


Figure \ref{fig_2} illustrates that all schemes achieve 100\% feasibility when the number of RIS elements is below 34. However, as the number of RIS elements increases beyond 34, the feasibility of all schemes rapidly decreases. Only the channel realizations with good channel conditions are feasible when the number of RIS elements is large.

As the CSI error increases, the feasibility of the proposed AO algorithm decreases. Figure \ref{fig_3} shows that the proposed AO algorithm is almost infeasible when the number of RIS elements is around 36. This is because when the CSI error is large, it is difficult to simultaneously satisfy the communication rate and the sensing CRB by only increasing the transmit power.

The trends of all curves in Fig. \ref{fig_4} are consistent. When the number of RIS elements is small (less than about 10), the transmit power decreases as the number of RIS elements increases. However, when the number of RIS elements is larger than about 10, the transmit power increases with the number of RIS elements. This is because when the number of RIS elements is small, the transmit power used to compensate for CSI error is also small. Nevertheless, when the number of RIS elements is large (less than about 50), the power loss caused by compensating for CSI error is greater than the power gain provided by the RIS, resulting in an increase of the transmit power as the number of RIS elements increases. When the number of RIS elements is relatively large (greater than about 50), the feasibility probability of the proposed algorithm is low and only channel realizations with good channel conditions are feasible. In such channel realizations, the transmit power required for robust transmission is low. Therefore, the transmission power gradually decreases as the number of RIS elements increases beyond 50. These trends in Fig. \ref{fig_4} break the intuitive conclusion that the more RIS elements, the greater the gain. When perfect CSI cannot be obtained, we should carefully adjust the number of RIS elements. Moreover, the transmit power obtained by RIS beamforming scheme 1 is slightly higher than the lower bound because the discrete phase shift projection introduces quantization errors, which degrade system performance. The transmit power obtained by RIS beamforming scheme 2 is higher than that obtained by RIS beamforming scheme 1 because RIS beamforming scheme 2 is a suboptimal algorithm with low computational complexity that iteratively solves the non-convex problem in (40a) using the gradient ascent method.

The transmit power of each scheme in Fig. \ref{fig_5} is higher than that in Fig. \ref{fig_4} because a larger transmit power is required to compensate for the increased CSI error. Moreover, the trends of the curves in Fig. \ref{fig_5} are consistent with those in Fig. \ref{fig_4}.

\begin{figure}[!t]
	\centering
	\includegraphics[width=0.45\textwidth]{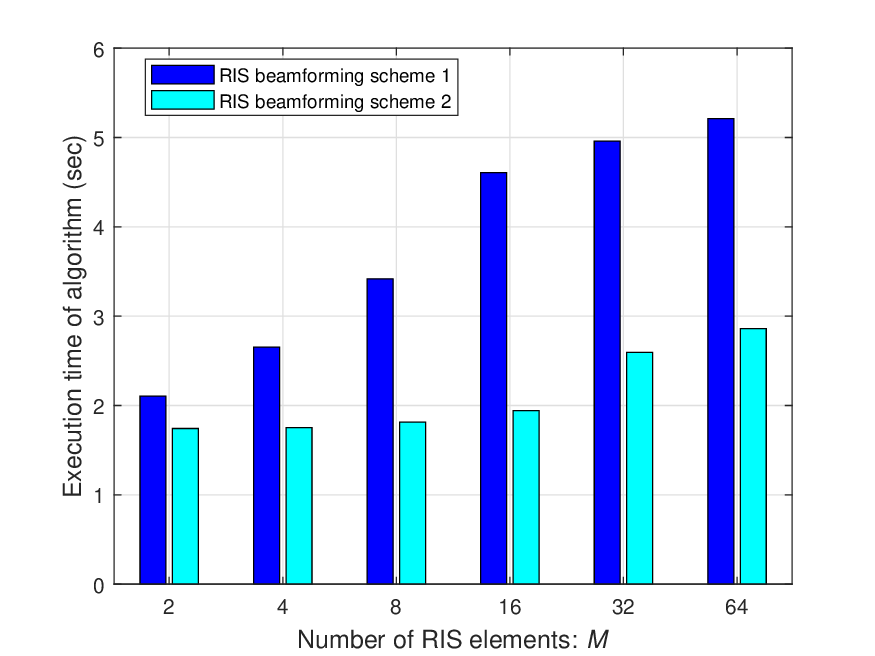}
	\caption{Execution time of algorithm versus the number of RIS elements when the number of RIS phase bits $d$ equals 4, $\gamma_{\text{BU},k}, \gamma_{\text{BRU},k}$, and $\varepsilon_l$ both equal 0.01.}
	\label{fig_6}
\end{figure}

Although the results in Fig. \ref{fig_4} and Fig. \ref{fig_5} illustrate that the transmit power of RIS beamforming scheme 1 is lower than that of RIS beamforming scheme 2, we have found in the previous section that the computational complexity of RIS beamforming scheme 2 is lower than that of RIS beamforming scheme 1. Figure \ref{fig_6} shows that as the number of RIS elements increases, the execution time difference between the two schemes becomes larger. When the number of RIS elements is large, such as 16, 32, and 64, the running time of RIS beamforming scheme 2 is 50\% shorter than that of scheme 1. 

\begin{figure}[!t]
	\centering
	\includegraphics[width=0.45\textwidth]{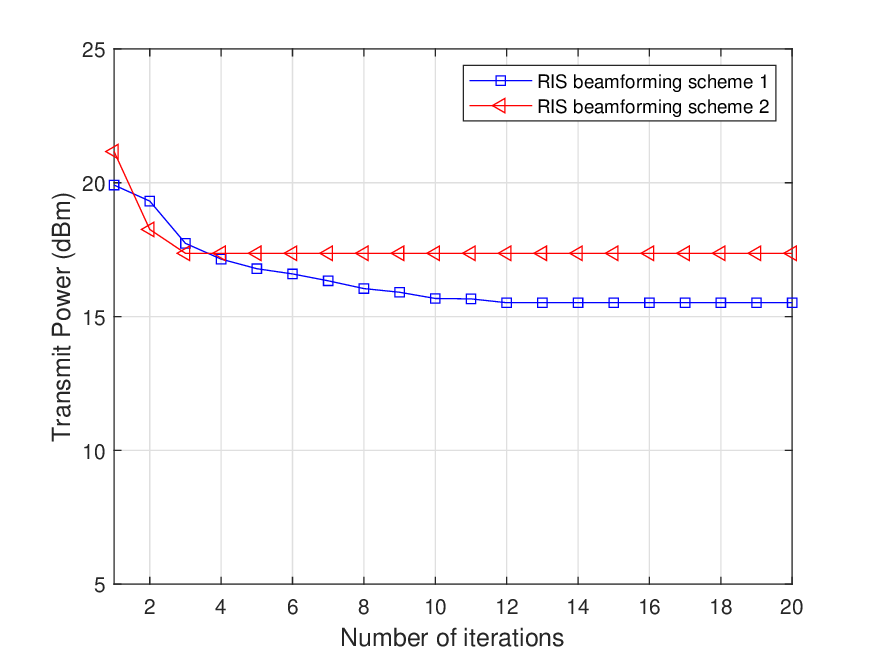}
	\caption{Transmit power versus the number of AO iterations when the number of RIS elements $M $is 32, the number of RIS phase bits $d$ equals 4, $\gamma_{\text{BU},k}, \gamma_{\text{BRU},k}$, and $\varepsilon_l$ both equal 0.01.}
	\label{fig_7}
\end{figure}

\begin{figure}[!t]
	\centering
	\includegraphics[width=0.45\textwidth]{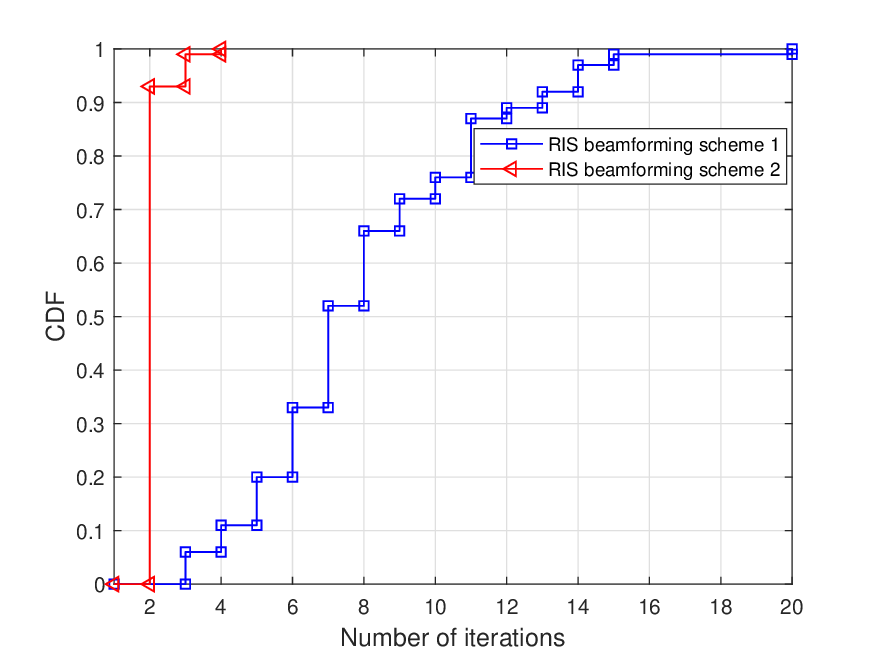}
	\caption{The cumulative distribution function (CDF) of the number of iterations.The number of RIS elements and BS antennas both equal to 16, the number of RIS phase bits $d$ equals 4, $\gamma_{\text{BU},k}, \gamma_{\text{BRU},k}$, and $\varepsilon_l$ both equal 0.01.}
	\label{fig_8}
\end{figure}

Figure \ref{fig_7} and Figure \ref{fig_8} depict the performance of the AO algorithms based on RIS beamforming scheme 1 and RIS beamforming scheme 2. All schemes achieve convergence within 20 AO iterations. Notably, the AO algorithm based on RIS beamforming scheme 1 achieves convergence in 20 iterations, while the AO algorithm based on RIS beamforming scheme 2 achieves convergence in only four iterations. RIS beamforming scheme 2 converges faster than RIS beamforming scheme 1 because the Gaussian randomization in the later scheme may lead to a degradation in system performance. Using a penalty function, RIS beamforming scheme 2 achieves faster convergence by directly iterating along the gradient direction within the feasible space.

\section{Conclusion}
This paper presents a novel approach for designing robust transmission schemes in RIS-assisted ISAC systems. A decomposition-based large deviation inequality approach is proposed to approximate non-convex outage probability constraints for communication and non-convex probability constraints for the sensing CRB. The AO algorithm is then employed to solve the joint beamforming problem. Specifically, the transmit beamforming problem is formulated as a SDP problem. We introduce two schemes to solve the discrete RIS beamforming problem, which are the SDR-based algorithm with high complexity and high precision and the SP-based GEMM algorithm with low complexity and low precision. Simulation results demonstrate the excellent performance of the proposed AO algorithms. It should be noted that increasing the number of RIS elements does not lead to better performance when the perfect CSI is not available. Thus, we should carefully select the number of RIS elements to guarantee communication rate and CRB for sensing parameter estimation.

\section*{Appendix A\\ Derivation of the CRB}
Assuming that the $T$ signal samples are statistically independent, the joint probability density can be written as
\begin{subequations}
	\label{eq100}
	\begin{align}
		&p_{\boldsymbol{y}_1,\cdots,\boldsymbol{y}_T\vert\boldsymbol{\phi}}\left(\boldsymbol{Y}\right)= \prod_{t=1}^T\frac{1}{\det\left(\pi \boldsymbol{K}_{\boldsymbol{Y}}\right)} \nonumber\\
		&\enspace\enspace\enspace\times\exp\left\{-\left(\boldsymbol{y}_t^H-\boldsymbol{m}_{\boldsymbol{Y}}^H\left(\boldsymbol{\phi}\right)\right)\boldsymbol{K}_{\boldsymbol{Y}}^{-1}\left(\boldsymbol{y}_t-\boldsymbol{m}_{\boldsymbol{Y}}\left(\boldsymbol{\phi}\right)\right)\right\},\tag{52}
	\end{align}
\end{subequations} 
where $\boldsymbol{K}_{\boldsymbol{Y}}$ is the covariance matrix for the received sensing signal, and $\boldsymbol{K}_{\boldsymbol{Y}}\triangleq \sigma_{\text{sen}}^2\textbf{I}$; $\boldsymbol{m}_{\boldsymbol{Y}}\left(\boldsymbol{\phi}\right)$ is the mean vector for the received sensing signal samples. Note that we neglect the subscript mark, i.e., sen, in $\boldsymbol{Y}_{\text{sen}}$ for simplicity.

The log-likelihood function is
\begin{subequations}
	\label{eq101}
	\begin{align}
		L_{\boldsymbol{Y}}(\boldsymbol{\phi})&=\ln p_{\boldsymbol{y}_1,\cdots,\boldsymbol{y}_T\vert\boldsymbol{\phi}}\left(\boldsymbol{Y}\right)\nonumber\\
		&=-T\ln \det(\boldsymbol{K}_{\boldsymbol{Y}})-TN\ln \pi\nonumber\\
		&\quad-\sum_{t=1}^{T}\left(\boldsymbol{y}_t^H-\boldsymbol{m}_{\boldsymbol{Y}}^H\left(\boldsymbol{\phi}\right)\right)\boldsymbol{K}_{\boldsymbol{Y}}^{-1}\left(\boldsymbol{y}_t-\boldsymbol{m}_{\boldsymbol{Y}}\left(\boldsymbol{\phi}\right)\right).\tag{53}
	\end{align}
\end{subequations}

We first derivate the CRB for single signal sample and then combine all the samples. The log-likelihood function for single sample can be expressed as
\begin{subequations}
	\label{eq102}
	\begin{align}
		&L_{\boldsymbol{y}}(\boldsymbol{\phi})=\ln p_{\boldsymbol{y}\vert\boldsymbol{\phi}}(\boldsymbol{y})\nonumber\\
		&=-\ln\det\left(\pi\boldsymbol{K}_{\boldsymbol{y}}\right)-\left\{\left[\boldsymbol{y}^H-\boldsymbol{m}_{\boldsymbol{y}}^H(\boldsymbol{\phi})\right]\boldsymbol{K}_{\boldsymbol{y}}^{-1}\left[\boldsymbol{y}-\boldsymbol{m}_{\boldsymbol{y}}(\boldsymbol{\phi})\right]\right\}.\tag{54}
	\end{align}
\end{subequations}

The Fisher's information for $\phi_l$ is defined as
\begin{equation}
	\label{eq103}
	f_l\triangleq-\text{E}\left[\frac{\partial^2L_{\boldsymbol{y}}(\boldsymbol{\phi})}{\partial\phi_l^2}\right],
\end{equation}
where the second derivative is expressed as
\begin{subequations}
	\label{eq104}
	\begin{align}
		\frac{\partial^2L_{\boldsymbol{y}}(\boldsymbol{\phi})}{\partial\phi_l^2}&=-2\text{Re}\bigg\{-\frac{\partial^2\boldsymbol{m}_{\boldsymbol{y}}^H(\boldsymbol{\phi})}{\partial\phi_l^2}\boldsymbol{K}_{\boldsymbol{y}}^{-1}\left[\boldsymbol{y}-\boldsymbol{m}_{\boldsymbol{y}}(\phi)\right] \nonumber\\
		&+\frac{\partial\boldsymbol{m}_{\boldsymbol{y}}^H(\phi)}{\partial\phi_l}\boldsymbol{K}_{\boldsymbol{y}}^{-1}\frac{\partial\boldsymbol{m}_{\boldsymbol{y}}(\phi)}{\partial\phi_l}\bigg\}.\tag{56}
	\end{align}
\end{subequations}
And the expectation of the second derivative is
\begin{equation}
	\label{eq105}
	\text{E}\left[\frac{\partial^2L_{\boldsymbol{y}}(\boldsymbol{\phi})}{\partial\phi_l^2}\right]=-2\text{Re}\left[\frac{\partial\boldsymbol{m}_{\boldsymbol{y}}^H(\phi)}{\partial\phi_l}\boldsymbol{K}_{\boldsymbol{y}}^{-1}\frac{\partial\boldsymbol{m}_{\boldsymbol{y}}(\phi)}{\partial\phi_l}\right].
\end{equation}
The CRB for the single sample can be expressed as
\begin{subequations}
	\label{eq106}
	\begin{align}
		\text{CRB}_l &\triangleq f_l^{-1}\nonumber\\
		&=\left\{2\text{Re}\left[\frac{\partial\boldsymbol{m}_{\boldsymbol{y}}^H(\phi)}{\partial\phi_l}\boldsymbol{K}_{\boldsymbol{y}}^{-1}\frac{\partial\boldsymbol{m}_{\boldsymbol{y}}(\phi)}{\partial\phi_l}\right]\right\}^{-1}. \tag{58}
	\end{align}
\end{subequations}
The CRB expression in (\ref{eq10}) can be obtained after we combine the CRBs from all samples.

\section*{Appendix B\\Proof of Proposition 1}
The second part of (\ref{eq17}) is rewritten as
\begin{subequations}
	\label{eq107}
	\begin{align}
		&2\text{Re}\Big[\gamma_{\text{BU},k}\left(\hat{\boldsymbol{h}}_{\text{BU},k}^H+\boldsymbol{\theta}^H\hat{\boldsymbol{H}}_{\text{BRU},k}\right)\boldsymbol{\Psi}_k\boldsymbol{e}_{\text{BU},k}\nonumber\\
		&\quad+\gamma_{\text{BRU},k}\left(\hat{\boldsymbol{h}}_{\text{BU},k}^H+\boldsymbol{\theta}^H\hat{\boldsymbol{H}}_{\text{BRU},k}\right)\boldsymbol{\Psi}_k\boldsymbol{E}_{\text{BRU},k}^H\boldsymbol{\theta}\Big]\nonumber\\
		&=2\text{Re}\Big[\gamma_{\text{BU},k}\left(\hat{\boldsymbol{h}}_{\text{BU},k}^H+\boldsymbol{\theta}^H\hat{\boldsymbol{H}}_{\text{BRU},k}\right)\boldsymbol{\Psi}_k\boldsymbol{e}_{\text{BU},k}\nonumber\\
		&+ \gamma_{\text{BRU},k}\text{vec}\left[\boldsymbol{\theta}\left(\hat{\boldsymbol{h}}_{\text{BU},k}^H+\boldsymbol{\theta}^H\hat{\boldsymbol{H}}_{\text{BRU},k}\right)\boldsymbol{\Psi}_k\right]^T\text{vec}\left(\boldsymbol{E}_{\text{BRU},k}\right)^*,\tag{59}
	\end{align}
\end{subequations}
where we user the property of the trace-operator and the vec-operator \cite[Eq. (521)]{ref47}, i.e., $\boldsymbol{a}^T\boldsymbol{Bc}=\text{Tr}(\boldsymbol{ac}^T\boldsymbol{B})=\text{vec}(\boldsymbol{ca}^T)^T\text{vec}(\boldsymbol{B})$. We can formulate (\ref{eq107}) as $2\text{Re}(\boldsymbol{t}_k^H\boldsymbol{e}_k)$ using the definition in (\ref{eq19}) and (\ref{eq21}).

The third part of (\ref{eq17}) is rewritten as
\begin{subequations}
	\label{eq108}
	\begin{align}
		&\Delta\boldsymbol{h}_{\text{BU},k}^H\boldsymbol{\Psi}_k\Delta\boldsymbol{h}_{\text{BU},k}+\boldsymbol{\theta}^H\Delta\boldsymbol{H}_{\text{BRU},k}\Delta\boldsymbol{H}_{\text{BRU},k}^H\boldsymbol{\theta} \nonumber\\
		&\quad+\Delta\boldsymbol{h}_{\text{BU},k}^H\boldsymbol{\Psi}_k\Delta\boldsymbol{H}_{\text{BRU},k}^H\boldsymbol{\theta}+\boldsymbol{\theta}^H\Delta\boldsymbol{H}_{\text{BRU},k}\boldsymbol{\Psi}_k\Delta\boldsymbol{h}_{\text{BU},k} \nonumber\\
		&=\gamma_{\text{BU},k}^2\boldsymbol{e}_{\text{BU},k}^H\boldsymbol{\Psi}_k\boldsymbol{e}_{\text{BU},k} \nonumber\\
		&\quad+\gamma_{\text{BU},k}\gamma_{\text{BRU},k}\boldsymbol{e}_{\text{BU},k}^H\left(\boldsymbol{\Psi}_k\otimes\boldsymbol{\theta}^T\right)\text{vec}\left(\boldsymbol{E}_{\text{BRU},k}\right)^*\nonumber\\
		&\quad+\gamma_{\text{BU},k}\gamma_{\text{BRU},k}\text{vec}\left(\boldsymbol{E}_{\text{BRU},k}\right)^T\left(\boldsymbol{\Psi}_k\otimes\boldsymbol{\theta}^*\right)\boldsymbol{e}_{\text{BU},k}\nonumber\\
		&\quad+\gamma_{\text{BRU},k}^2\text{vec}\left(\boldsymbol{E}_{\text{BRU},k}\right)^T\left(\boldsymbol{\Phi}\otimes\left(\boldsymbol{\theta}\boldsymbol{\theta}^H\right)\right)\text{vec}\left(\boldsymbol{E}_{\text{BRU},k}\right)^*,\tag{60}
	\end{align}
\end{subequations}
where we use the property of the vec-operator \cite[Eq. (524)]{ref47}, i.e., $\boldsymbol{a}^T\boldsymbol{XBX}^T\boldsymbol{c}=\text{vec}(\boldsymbol{X})^T\left(\boldsymbol{B}\otimes\boldsymbol{ca}^T\right)\text{vec}(\boldsymbol{X})$. Please refer to B.1.1 of \cite{ref47} for a proof for this property. We can formulate (\ref{eq108}) as $\boldsymbol{e}_k^H\boldsymbol{Q}_k\boldsymbol{e}_k$ using the definition in (\ref{eq19}) and (\ref{eq20}).

The first part and the last part of (\ref{eq17}) form $s_k$. Thus, we obtain (\ref{eq18}). This completes the proof of Proposition 1.

\section*{Appendix C\\Proof of Proposition 2}
In order to derive a convex restriction of (\ref{eq18}) using Lemma 1, we set $\boldsymbol{x}=\boldsymbol{e}_k$, $\boldsymbol{Q}=\boldsymbol{Q}_k$, $\boldsymbol{r}=\boldsymbol{r}_k$, $\eta_k=\text{Tr}(\boldsymbol{Q}_k)+s_k$, and $T_k=v_k\Vert\boldsymbol{Q}_k\Vert_F+\frac{1}{\sqrt{2}}\Vert\boldsymbol{r}_k\Vert$. We choose $v_k$ as the solution to $[1-1/(2v_k^2)]v_k=\sqrt{\ln(1/\rho_k)}$. We could observe that when $\eta_k\textgreater2\bar{\theta}_kv_kT_k$, i.e., $\eta_k\ge2\sqrt{\ln(1/\rho_k)}T_k$, the outage probability constraint (\ref{eq18}) is always satisfied. It is shown as the following equation.
\begin{subequations}
	\label{eq109}
	\begin{align}
		&\text{Pr}\left\{\boldsymbol{e}_k^H\boldsymbol{Q}_k\boldsymbol{e}_k+2\text{Re}(\boldsymbol{r}_k^H\boldsymbol{e}_k)+s_k\leq0\right\}\nonumber\\
		&=\exp\left[-\frac{\bar{\theta}_kv_k\eta_k}{T_k}+\left(\bar{\theta}_kv_k\right)^2\right]\textless\exp\left[-\left(\bar{\theta}_kv_k\right)^2\right]=\rho_k.\tag{61}
	\end{align}
\end{subequations}
Further, (\ref{eq18}) is also satisfied if $\eta_k=2\bar{\theta}_kv_kT_k$. However, (\ref{eq18}) may not be satisfied if $\eta_k\textless2\bar{\theta}_kv_kT_k$. Thus, we add constraint for this situation to ensure the outage probability is no more that $\rho_k$. The constraint is expressed as
\begin{equation}
	\label{eq110}
	\exp\left\{-\frac{\left[\text{Tr}(\boldsymbol{Q}_k)+s_k\right]^2}{4T_k^2}\right\}\leq\rho_k.
\end{equation}
Then, we can obtain the SOC constraint (\ref{eq23}). This completes the proof of Proposition 2.

\section*{Appendix D\\Proof of Proposition 3}
In order to derive a convex restriction of (\ref{eq28}) using Lemma 1, we set $\boldsymbol{x}=e_{\text{sen},l}$, $\boldsymbol{Q}=\tilde{q}_l$, $\boldsymbol{r}=\tilde{r}_l$, $\eta=\tilde{q}_l+\tilde{s}_l$, and $\tilde{T}_l=\tilde{v}_l\tilde{q}_l+\frac{1}{\sqrt{2}}\tilde{r}_l$. We omit the intermediate steps since they are similar with Appendix C. The convex constraint is expressed as
\begin{equation}
	\label{eq111}
	\exp\left\{-\frac{\left[\tilde{q}_l+\tilde{s}_l\right]^2}{4\tilde{T}_l^2}\right\}\leq p_l.
\end{equation}
Then, we can obtain the SOC constraint (\ref{eq29}). This completes the proof of Proposition 3.

\begin{IEEEbiography}[{\includegraphics[width=1in,height=1.25in,clip,keepaspectratio]{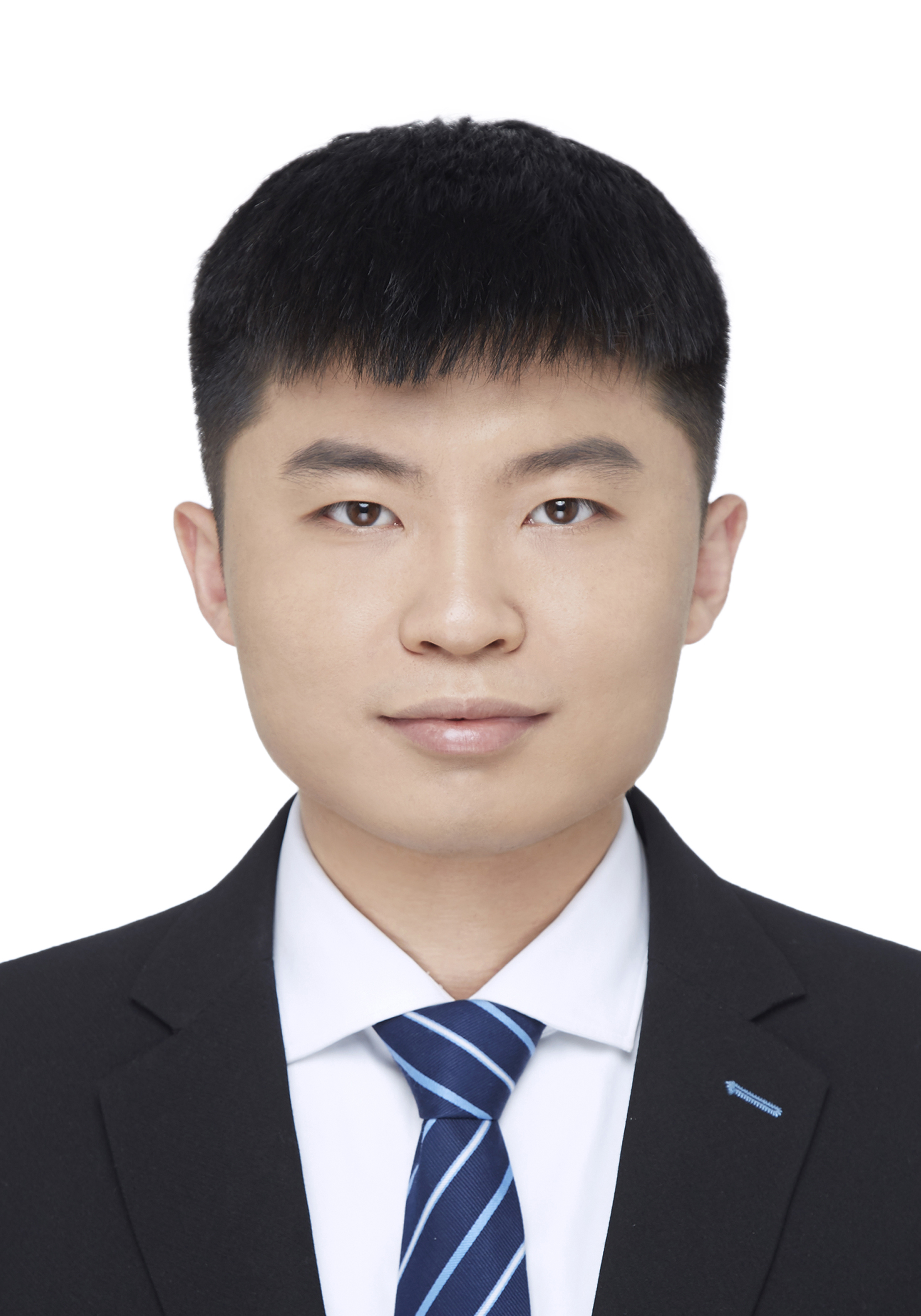}}]{Yongqing Xu}
	(Graduate Student Member, IEEE) received the bachelor's degree in communication engineering from the Xi'an University of Posts and Telecommunications (XUPT), Xi'an, China, in 2019. He is currently pursuing the Ph.D. degree with the School of Information and Communication Engineering of Beijing University of Posts and Telecommunications (BUPT), Beijing, China. His research mainly focuses on integrated sensing and communication (ISAC), reconfigurable intelligent surface (RIS), and the convex optimization.
\end{IEEEbiography}

\begin{IEEEbiography}[{\includegraphics[width=1in,height=1.25in,clip,keepaspectratio]{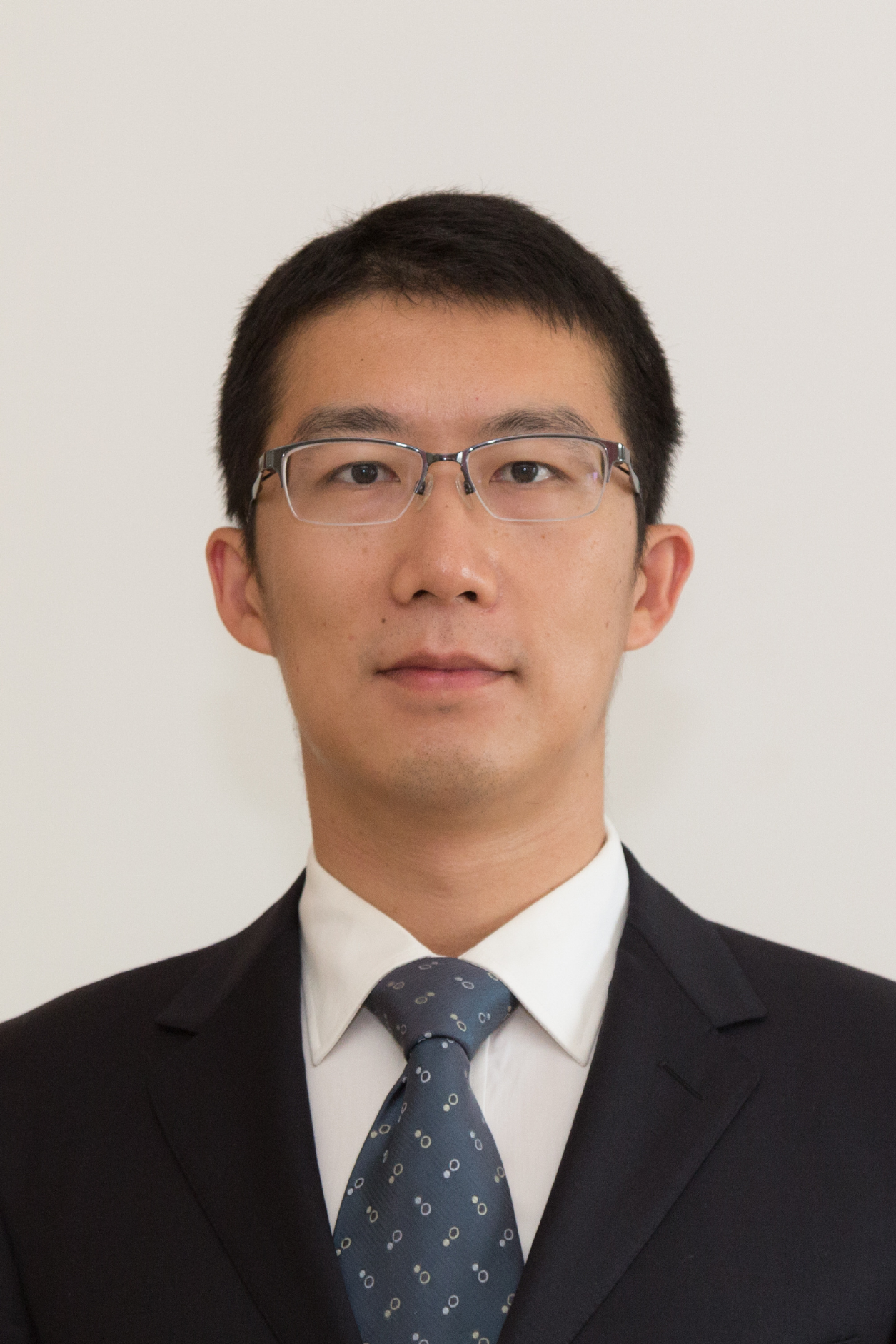}}]{Yong Li}
	(Member, IEEE) received the Ph.D. degree in signal and information processing from Beijing University of Posts and Telecommunications (BUPT), Beijing, China, in 2009.
	
	He is currently a Full Professor with the School of Information and Communication Engineering, BUPT. He has published more than 100 papers in journals, conference proceedings, and workshops. He holds more than 60 patents, including two U.S. patents. His current research interests include integrated sensing and communication, space-air-ground integrated networks, beyond-5G systems, and over-the-air (OTA) testing.
	
	Dr. Li was a recipient of the First Grade Award of Technological Invention from the China Institute of Communications.
\end{IEEEbiography}

\begin{IEEEbiography}[{\includegraphics[width=1in,height=1.25in,clip,keepaspectratio]{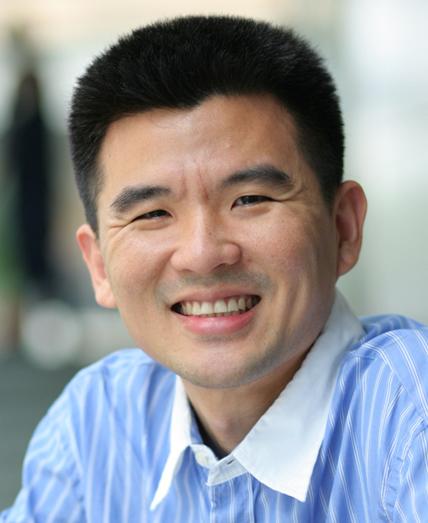}}]{Tony Q.S. Quek}
	(S'98-M'08-SM'12-F'18) received the B.E.\ and M.E.\ degrees in electrical and electronics engineering from the Tokyo Institute of Technology in 1998 and 2000, respectively, and the Ph.D.\ degree in electrical engineering and computer science from the Massachusetts Institute of Technology in 2008. Currently, he is the Cheng Tsang Man Chair Professor with Singapore University of Technology and Design (SUTD) and ST Engineering Distinguished Professor. He also serves as the Director of the Future Communications R\&D Programme, the Head of ISTD Pillar, and the Deputy Director of the SUTD-ZJU IDEA. His current research topics include wireless communications and networking, network intelligence, non-terrestrial networks, open radio access network, and 6G.
	
	Dr.\ Quek has been actively involved in organizing and chairing sessions, and has served as a member of the Technical Program Committee as well as symposium chairs in a number of international conferences. He is currently serving as an Area Editor for the {\scshape IEEE Transactions on Wireless Communications}. 
	
	Dr.\ Quek was honored with the 2008 Philip Yeo Prize for Outstanding Achievement in Research, the 2012 IEEE William R. Bennett Prize, the 2015 SUTD Outstanding Education Awards -- Excellence in Research, the 2016 IEEE Signal Processing Society Young Author Best Paper Award, the 2017 CTTC Early Achievement Award, the 2017 IEEE ComSoc AP Outstanding Paper Award, the 2020 IEEE Communications Society Young Author Best Paper Award, the 2020 IEEE Stephen O. Rice Prize, the 2020 Nokia Visiting Professor, and the 2022 IEEE Signal Processing Society Best Paper Award. He is an IEEE Fellow, a WWRF Fellow, and a Fellow of the Academy of Engineering Singapore.
\end{IEEEbiography}

\end{document}